\documentclass[fleqn,usenatbib]{mnras}

\usepackage[T1]{fontenc}

\DeclareRobustCommand{\VAN}[3]{#2}
\let\VANthebibliography\thebibliography
\def\thebibliography{\DeclareRobustCommand{\VAN}[3]{##3}\VANthebibliography}

\usepackage{graphicx}	% Including figure files
\usepackage{amsmath}	% Advanced maths commands
\usepackage{amssymb}	% Extra maths symbols
\usepackage{upgreek}
\usepackage{newtxtext,newtxmath}

\hypersetup{citecolor=blue}
\newcommand{\angstrom}{\mbox{\normalfont\AA}}
\newcommand\solarmass{\ensuremath{\mathrm{M}_{\odot}}}

\title[Triage of \textit{Gaia} astrometric binaries]
{Triage of the \textit{Gaia} DR3 astrometric orbits. II. A census of white dwarfs}

\author[S. Shahaf et al.]{
S. Shahaf,$^{1}$\thanks{These authors contributed equally.}\thanks{E-mail: \href{mailto:sahar.shahaf@weizmann.ac.il}{sahar.shahaf@weizmann.ac.il}}
N. Hallakoun,$^{1}$\footnotemark[1]\thanks{E-mail: \href{mailto:sahar.shahaf@weizmann.ac.il}{naama.hallakoun@weizmann.ac.il}}
T. Mazeh,$^{2}$
S. Ben-Ami,$^{1}$
P. Rekhi,$^{1}$,
K. El-Badry$^{3,4}$ and
S. Toonen$^{5}$
\\
$^{1}$Department of Particle Physics and Astrophysics, Weizmann Institute of Science, Rehovot 7610001, Israel\\
$^{2}$School of Physics and Astronomy, Tel Aviv University, Tel Aviv, 6997801, Israel\\
$^{3}$Center for Astrophysics $|$ Harvard \& Smithsonian, 60 Garden Street, Cambridge, MA 02138, USA\\
$^{4}$Department of Astronomy, California Institute of Technology, Pasadena, CA 91125, USA\\
$^{5}$Astronomical Institute Anton Pannekoek, University of Amsterdam, Science Park 904, 1098 XH Amsterdam, The Netherlands\\
}
\date{Accepted XXX. Received YYY; in original form ZZZ}

\pubyear{2023}

\begin{document}
\label{firstpage}
\pagerange{\pageref{firstpage}--\pageref{lastpage}}
\maketitle

% Abstract of the paper
\begin{abstract}
The third data release of \textit{Gaia} was the first to include orbital solutions assuming non-single stars. Here, we apply the astrometric triage technique of Shahaf et al. to identify binary star systems with companions that are not single main-sequence stars. \textit{Gaia}'s synthetic photometry of these binaries is used to distinguish between systems likely to have white-dwarf companions and those that may be hierarchical triples. The study uncovered a population of nearly $3\,200$ binaries, characterised by orbital separations on the order of an astronomical unit, in which the faint astrometric companion is probably a white dwarf. This sample increases the number of orbitally solved binary systems of this type by about two orders of magnitude.
Remarkably, over $110$ of these systems exhibit significant ultraviolet excess flux, confirming this classification and, in some cases, indicating their relatively young cooling ages. We show that the sample is not currently represented in synthetic binary populations, and is not easily reproduced by available binary population synthesis codes. Therefore, it challenges current binary evolution models, offering a unique opportunity to gain insights into the processes governing white-dwarf formation, binary evolution, and mass transfer.
\end{abstract}

% Select between one and six entries from the list of approved keywords.
% Don't make up new ones. 

%%%%%%%%%%%%%%%%%%%%%%%%%%%%%%%%%%%%%%%%%%%%%%%%%%

%%%%%%%%%%%%%%%%% BODY OF PAPER %%%%%%%%%%%%%%%%%%
\defcitealias{shahaf23}{Paper I}
\defcitealias{arenou22}{Gaia Collaboration, Arenou,
et al. 2022}
% ===============================================
%               1) Introduction
% ===============================================
\section{Introduction}
The detection of white dwarfs (WDs) in wide binaries goes back to the early days of modern astrometry, when in \citeyear{bessel1844} Friedrich Bessel reported on peculiar changes in the sky position of Sirius and Procyon. After considering several possibilities, Bessel concluded that Sirius and Procyon are not single stars and move under the gravitational force induced by unseen companions of unknown nature. The proximity of Sirius and Procyon to Earth eventually allowed for direct imaging of these faint companions \citep[][]{bond1862,schaeberle1896}, and led to their identification as WDs.

The detection of WDs in binaries that are located at distances considerably greater than that of Sirius and Procyon relies on the indirect method initially proposed by Bessel---detecting the faint WD through its effect on the orbital motion of its visibly brighter, usually main-sequence (MS), primary star. However, the analysis of the binary motion alone is often insufficient to determine the nature of the faint companion. Traditionally, identifying WD companions has depended on observing their contributions to the system's brightness in short-wavelength bands \citep[e.g.][and references therein]{pearsons16}. The flux ratio between the WD companions and their bright stars can sometimes be detected in those bands, as the WDs can be much hotter than their companions.

Still, in many cases, the light contribution of the WD is too small to be detected, even in the short-wavelength bands. Therefore, it is likely that many WD companions to MS stars eluded detection or were incorrectly classified \citep[][]{holberg13}, particularly if their MS companion is of spectral type earlier than M0.

This situation has changed drastically with the release of a catalogue of nearly $170\,000$ astrometric binaries, many of which are presumed to have unseen WD companions, in the third data release (DR3) of \textit{Gaia} \citepalias{arenou22}. 
Compact companions in such systems can be identified by ruling out all other possibilities. These include a relatively faint single MS companion or, in case the observed binary is wide enough, a companion that is by itself a short-period binary consisting of two faint MS stars \citep[e.g.][]{shahaf19, shenar22, janssens22, shahaf23, gaiaBH1, Chakrabarti2023}. 
Obtaining a large sample of binaries with WD secondaries may refine our understanding of the late stages of stellar evolution of the binary population, the WD initial-to-final mass relation, and the processes governing mass transfer and co-evolution in binary systems (\citealt{gratton21, venner23, escorza23, zhang23, sayeed23}). 

The first paper of this series, \citet[Paper I henceforth]{shahaf23}, identified the astrometric binaries that are most unlikely to have single MS companions. Based on this classification, the work presented here divides those binaries into two sub-groups: those whose companions are likely to be short-period faint MS binaries by themselves (i.e., hierarchical triple systems) and those that probably have WD companions. The division between these two sub-groups is done by identifying the light contribution of the faint companion(s). Instead of searching for the short-wavelength contribution from the WD, we utilise the availability of \textit{Gaia}'s synthetic photometry to identify the long-wavelength contribution of faint MS close-binary companions. We assume that the binaries with no long-wavelength excess are the ones with WD companions.

The structure of this paper is as follows: In Section~\ref{sec: sample selection}, we outline the process we used to select our sample and provide a review of its characteristics. Section~\ref{sec: wds} discusses the WD population within our sample and presents our findings. In particular, subsection~\ref{sec: uv} is dedicated to presenting the ultraviolet (UV) emission detected in some of our targets. In Section~\ref{sec:biases}, we discuss the various biases and selection effects in our sample. Finally, in Section~\ref{sec: summary}, we summarise our findings and discuss potential avenues for further research.

% --------------------
% --------------------
\section{Sample selection}
\label{sec: sample selection}
% --------------------
% --------------------

\begin{figure}
    \centering
    \includegraphics[width=0.99\columnwidth]{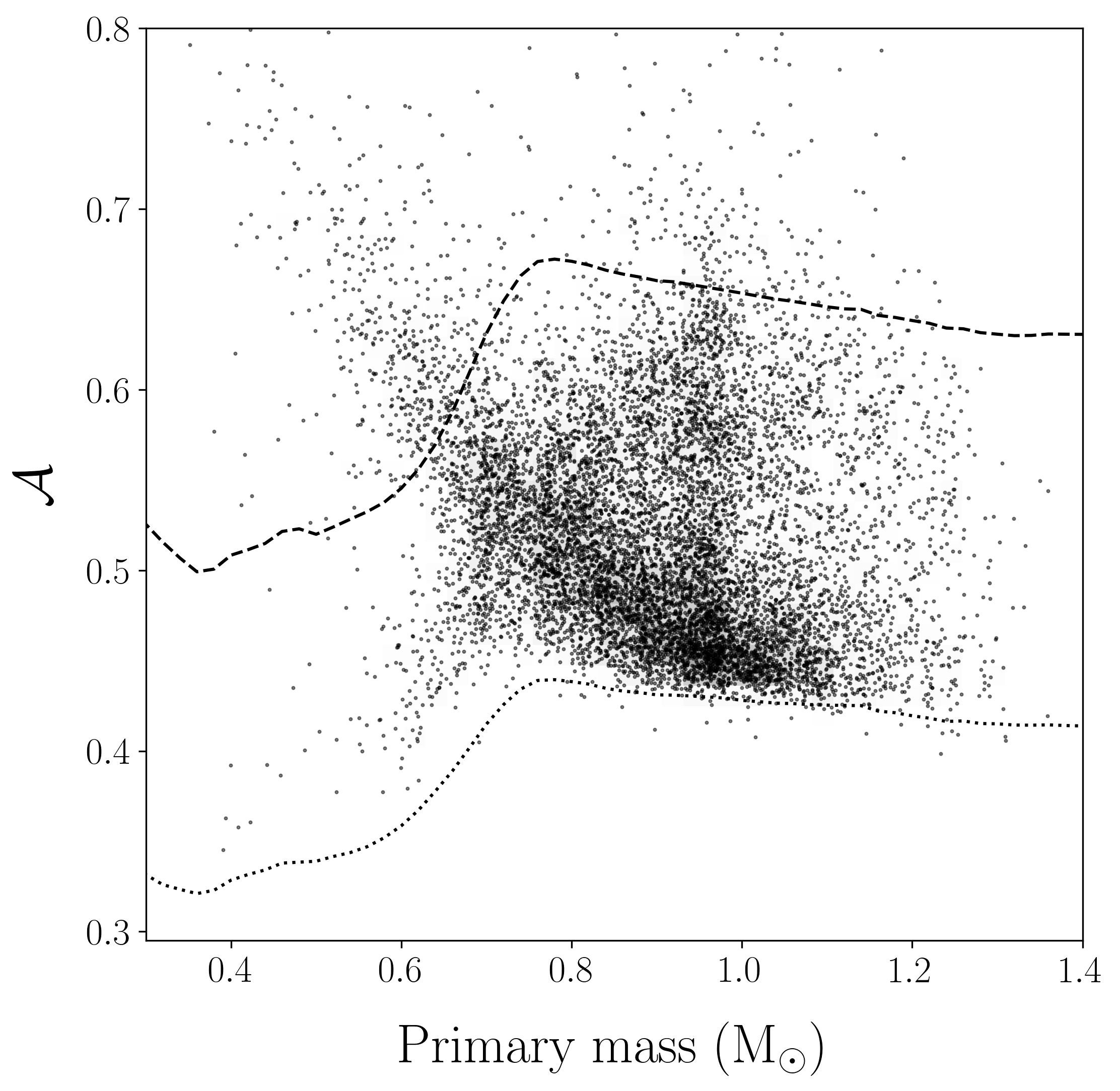}
    \caption{The astrometric mass-ratio function (AMRF) versus the primary mass of the $9\,786$ selected non-\textit{class-I} systems. The dotted line separates between \textit{class-I} and \textit{class-II} regions, and the dashed line separates between \textit{class-II} and \textit{-III}.}
    \label{fig:AMRF_vs_m1_all}
\end{figure}

\begin{figure*}
    \begin{minipage}{0.475\textwidth}
    \centering
        \includegraphics[width=0.95\columnwidth]{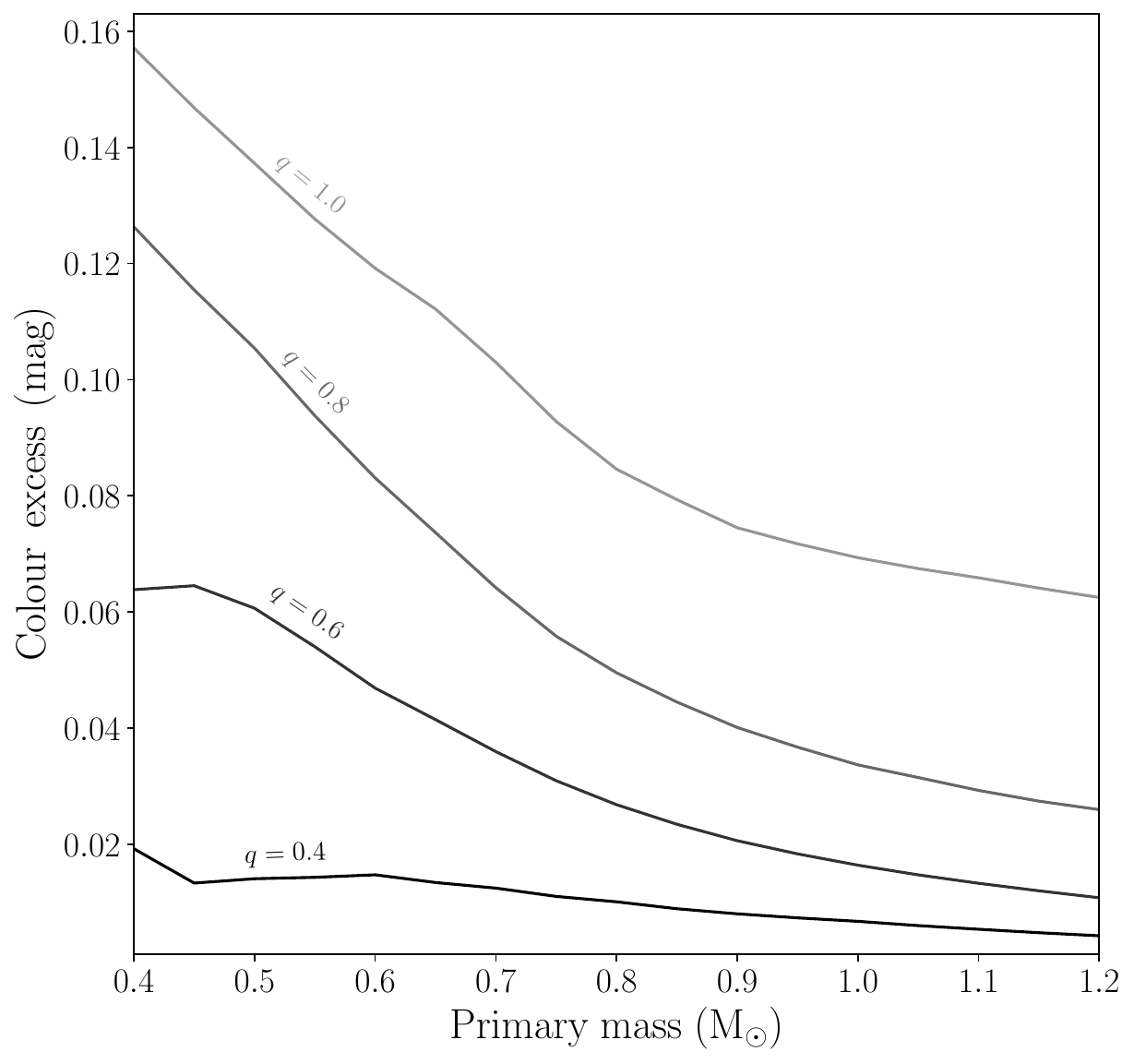} 
        \caption{The expected $B-I$ colour excess induced by an unresolved binary system versus the mass of the astrometric primary star. Each line in the diagram corresponds to a different mass ratio, $q$, between the astrometric primary and secondary. The close binary is assumed to consist of two equal-mass MS stars.}
        \label{fig: triple color excess}
    \end{minipage}\hfill
    \begin{minipage}{0.475\textwidth}
        \centering
        \includegraphics[width=0.95\columnwidth]{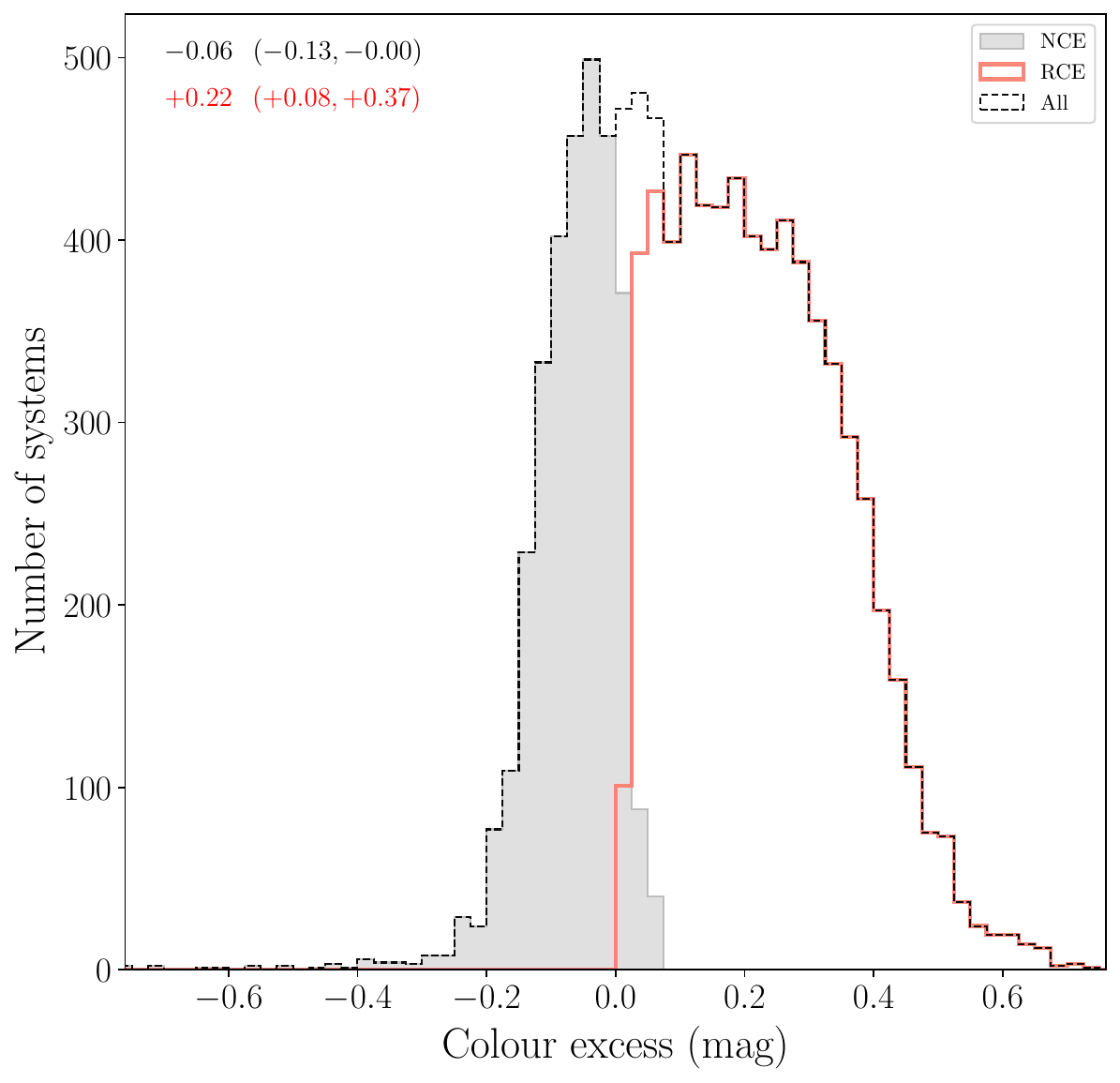}
        \caption{The estimated $B-I$ colour excess distribution of the red-colour excess (RCE) and no-colour excess (NCE) populations. The black dashed line shows the distribution of the full sample. The colour excess median and $36 - 84$th percentile range for the NCE (top row) and RCE (bottom row) populations appear in the top-left corner.}
    \label{fig: color hist}
    \end{minipage}
\end{figure*}

To identify astrometric binaries with WD companions, we compiled a sample of systems where the secondary component is less likely to be a single MS star. We then utilised the photometric measurements provided by\textit{Gaia} to identify excess near-infrared emission to distinguish between hierarchical triple systems and MS+WD binaries. The data underlying this analysis are obtained from table~1 of \citetalias{shahaf23} and the following \textit{Gaia} DR3 tables: \texttt{gaia\_source},  \texttt{nss\_two\_body\_orbit}, \texttt{binary\_masses}, and \texttt{synthetic\_photometry\_gspc}. 

% ----------------------------
\subsection{Astrometric triage}
% ----------------------------

The \textit{astrometric mass-ratio function} (AMRF; \citealt{shahaf19}) is defined as
%---------------------------------------
\begin{equation}
    \mathcal{A} \equiv \frac{\alpha_0}{\varpi} \bigg(\frac{M_1}{\textrm{M}_\odot}\bigg)^{-1/3} \bigg(\frac{P}{\textrm{yr}}\bigg)^{-2/3}\, , 
    \label{eq: AMRF 1}
\end{equation}
%--------------------------------------
where $P$ is the orbital period, $M_1$ is the primary mass, $\varpi$ is the parallax,  and $\alpha_0$ is the angular semimajor axis of the photocentric orbit.

\citet{shahaf19} showed that based on the AMRF value, one can assign each binary with an MS primary to one of three classes: \textit{class-I}, where the companion is most likely a single MS star; \textit{class-II}, where the companion cannot be a single MS star, but can still be a short-period binary of two MS stars; and, \textit{class-III}, where the companion cannot be a single MS star or a close MS binary; therefore the faint astrometric secondary in such systems is probably a compact object.

\citetalias{shahaf23} used the AMRF approach to analyse a sample of $101\,380$ astrometric binaries reported in \textit{Gaia} DR3. The selected binaries have MS primary stars, orbital periods shorter than $1\,000$ days, and orbital solutions that passed several quality criteria (see therein), estimating the probability of each system to be a \textit{class-II} or \textit{class-III} astrometric binary. These probabilities, denoted Pr~II and Pr~III, are provided in table~1 therein. 
\citetalias{shahaf23} used these classification probabilities to compile a short list of $177$ systems for which Pr~III is close to unity. Assuming that \textit{Gaia}'s orbital parameters are valid, their faint astrometric secondary components are likely to be compact objects. In this work, we take a different approach, excluding \textit{class-I} systems instead of looking for \textit{class-III} systems. Namely, we consider astrometric orbits for which the secondary is probably not a single MS star. 

The \textit{class-I} probability, Pr~I, was not provided in \citetalias{shahaf23} but can be calculated using the other two classification probabilities,
\begin{equation}
    {\rm Pr \, I} = 1 - {\rm Pr \, II} - {\rm Pr \, III} + \frac{2}{N+1},
\end{equation}
where $N=10^5$. We include in our sample only systems that satisfy the qualifying conditions outlined in \citetalias{shahaf23} (see section 3.1, therein) and have Pr~I below $10$~per cent. This selection yielded $11\,190$ non-\textit{class-I} systems. 

Binaries with a primary mass larger than ${\sim}1.2$\,${\rm M}_\odot$ (within $1\sigma$), for which WD companions are unlikely to be classified as \textit{class-II} or \textit{-III} systems (see figure 2 of \citetalias{shahaf23}), were excluded from our sample. This selection criterion resulted in $10\,080$ binaries. We then removed additional binaries for which the \textit{Gaia} Synthetic Photometry Catalogue \citep[GSPC;][]{montegriffo22} did not provide estimates in the Johnson-Kron-Cousins bands (JKC; see below). Following this final step, we were left with a sample of $9\,786$ astrometric binaries, presented in Fig.~\ref{fig:AMRF_vs_m1_all}.

% ------------------------
\subsection{Colour excess}
% ------------------------
\label{sec: colour excess}

\subsubsection{Expected colour excess}
We use \textit{Gaia}'s spectrophotometric measurements to distinguish between hierarchical triples and MS+WD binaries and resolve this classification ambiguity. Fig.~\ref{fig: triple color excess} shows the expected colour excess induced on the observed MS primary by an unresolved close-binary companion, using JKC's $B$ and $I$ bands. We define the colour excess as the difference between the triple system's and the primary star's colours.  Formally, this is given by
\begin{equation}
\begin{split}
    \Delta\left({B-I}\right) =& -2.5 \log_{10} \left( 10^{-0.4B_1} + 10^{-0.4B_{2\text{a}}} + 10^{-0.4B_{2\text{b}}} \right) \\
    & + 2.5 \log_{10} \left( 10^{-0.4I_1} + 10^{-0.4I_{2\text{a}}} + 10^{-0.4I_{2\text{b}}} \right) \\
    & - \left(B_1-I_1\right)\,,
\end{split}
\end{equation}
where the subscript $1$ refers to the astrometric primary. The other two subscripts, 2a and 2b, refer to the two components of the astrometric secondary.

The estimates in Fig.~\ref{fig: triple color excess} were calculated using PARSEC\footnote{We use PARSEC (PAdova and tRieste Stellar Evolution Code) version 1.2S, available online via \url{http://stev.oapd.inaf.it/cmd}.} isochrones \citep[]{Bressan_2012, Chen_2014, Chen_2015, Tang_2014}. The close binary is assumed to comprise two equal-mass MS stars since this configuration yields the largest mass-luminosity ratio between the astrometric components. The figure illustrates that per-cent level photometric precision should be sufficient to identify the close-binary companions for systems with primaries less massive than ${\sim}1.2 \,\solarmass$. 

\subsubsection{Observed colour excess}
\textit{Gaia} DR3 catalogue provides flux-calibrated low-resolution spectrophotometric measurements in the wavelength range $330${--}$1050$\,nm. Based on these data, the \textit{Gaia} Synthetic Photometry Catalogue \citep[GSPC;][]{montegriffo22} published synthetic photometric measurements with accuracy as low as ${\sim}10$\,mmag. We compared the location of each source on the \textit{Gaia} colour-magnitude diagram (CMD) to its theoretically expected position, assuming it is a single star, in search of excess infrared flux. 

The system's mass, age, metallicity, and extinction along the line of sight determine the theoretical CMD position. As presented below, we calculated the theoretical position assuming a fixed age of 2 Gyr. Each system's mass, metallicity and extinction are accounted for, along with their estimated uncertainties. This theoretical position is compared with the observations to identify and quantify the infrared excess flux.

We used \citet{Zhang_2023} metallicity estimates. These values were derived in a forward-modelling approach, using \textit{Gaia}'s flux-calibrated low-resolution spectrophotometric measurements (i.e., \textit{Gaia}'s blue photometre (BP)/red photometre (RP) spectra, or ``XP spectra'' for short). The data-driven model \citet{Zhang_2023} developed was trained on atmospheric parameters reported by the Large Sky Area Multi-Object Fibre Spectroscopic Telescope (LAMOST) survey. Using its XP spectrum, the model provides a given target's metallicity, surface gravity, and effective temperature. The effects of binarity were not included in the model and may bias the estimated values of binaries with flux ratios of order unity \citet{Zhang_2023}. However, considering the typically extreme flux ratios of the systems discussed in this work, we expect this issue to have a limited impact (but see the discussion in Section~\ref{sec:biases}).

Systems flagged to have uncertain metallicity estimates by \citet[][their $\texttt{quality\_flags}~{\geq}~8$]{Zhang_2023} were assigned with [M/H]$~{=}~0.00 \pm 0.25$ dex. This choice is justified by the age-metallicity relation of stars in the solar neighbourhood \citep{RebassaMansergas_2021}. Systems with metallicity estimates laying outside of the PARSEC isochrones range ($-2\leq [\text{M/H}] \leq 0.6$) were assigned with the corresponding limiting values. 

We used publicly available dust maps to account for reddening and extinction. \citet{Green_2019} have created  \texttt{Bayestar19}: a three-dimensional dust map that covers the northern hemisphere down to a declination of $-30$\,deg. We used the \texttt{dustmaps} \texttt{Python} package to obtain the 16\textsuperscript{th}, 50\textsuperscript{th}, and 84\textsuperscript{th} percentiles of \texttt{Bayestar19} extinction. These values were then multiplied by 0.884 to obtain the corresponding $E(B-V)$ values.\footnote{See \href{http://argonaut.skymaps.info/usage\#units}{argonaut.skymaps.info/usage}.} 
For the southern systems, which were not included in \texttt{Bayestar19}, we used the three-dimensional dust map of \citet{Lallement_2019}.\footnote{\url{https://astro.acri-st.fr/gaia_dev/}} This map does not include error estimates. Considering systems included in both maps, the median difference between \texttt{Bayestar19} and the \citet{Lallement_2019} extinction is 0.02\,mag. We therefore used this value as the uncertainty for the $E(B-V)$ values of the southern hemisphere systems. Finally, we converted the $E(B-V)$ values to $E(B-I)$ and $A_V$, using the relations 
\begin{equation}
\begin{aligned}
   E(B-I) = \,\,& (R_B-R_I)\times E(B-V) \,\,\,\text{and} \\[8pt]
   A_V    = \,\,& R_V \times  E(B-V) \,,
\end{aligned}
\end{equation}
where the coefficients $R_B$, $R_I$ and $R_V$ were taken as $3.626$, $1.505$ and $3.1$, respectively \citep[according to][]{Munari_1996, Schlafly_2011}.

We estimated the colour excess using the \texttt{Python} package \texttt{stam}\footnote{Available online via \href{https://github.com/naamach/stam}{github.com/naamach/stam}.} \citep{hallakoun21}: Using PARSEC main-sequence isochrones, we generated a two-dimensional interpolant grid\footnote{Created using \texttt{scipy}'s linear radial basis function interpolation, documented at \href{https://docs.scipy.org/doc/scipy/reference/generated/scipy.interpolate.Rbf.html}{docs.scipy.org/doc/scipy}.} representing the relation between the $B-I$ colour index, the absolute $V$-band magnitude, and the metallicity. We limited the maximal mass of the isochrones to 1.17\,\solarmass\ to avoid tracks that start to evolve off the MS, and assumed a fixed age of 2\,Gyr. Stars do not significantly change their mass or appearance during their MS lifetime; therefore, this choice of isochrone age does not severely affect colour excess estimates. Our specific stellar age choice is consistent with the ages estimated for the more massive stars in our sample and justified by the observed evidence for a star formation burst in the Milky Way's disc about 2\,Gyr ago \citep{Mor_2019}. 

The observed colour excess is defined as
\begin{equation}
\label{eq: ce}
    \Delta \left(B-I\right) = \left(B-I\right)_\text{observed} - \left(B-I\right)_\text{expected},
\end{equation}
where $\left(B-I\right)_\text{observed}$ is the observed dereddened $B-I$ colour index, and the $\left(B-I\right)_\text{expected}$ is the theoretically expected $B-I$ value, assuming it is a single MS star. We calculated the theoretically expected values using the interpolant grid described above, with the system's estimated metallicity and extinction-corrected absolute $V$-band magnitude as its independent variables.

We conduct a Monte-Carlo experiment to account for the uncertainties in the observed quantities and provide confidence intervals for the estimated colour excess. For each astrometric binary, we generated $10^4$ random realisations of its position on the metallicity-absolute magnitude grid. The absolute $V$-band magnitude of the system is not expected to change by more than $\sim 0.1$\,mag by the presence of a WD companion (compared to a single MS star) for almost all our sample, except for those with primaries less massive than $\sim 0.3$\,\solarmass\ and warm WD companions (see Section~\ref{sec:biases} and Fig.~\ref{fig:PhotometricExcess_G} below). We assumed that the metallicity and the absolute magnitude follow an uncorrelated bi-variate Gaussian distribution. The measurement uncertainties of the observed colour index and absolute magnitude were calculated following \citet{hallakoun21}. For each instance, we calculated $\Delta \left(B-I\right)$ as described in equation~(\ref{eq: ce}). This sampled distribution's mean and standard deviation provide our estimates for the colour excess expected value and uncertainty. 
We provide the colour-excess estimates for all the targets in our sample in Table~\ref{tab: table1}.

The colour excess median uncertainty (first percentile) value is $0.15$ ($0.09$) mag. A systematic deviation of around $0.1$ dex to our metallicity estimates will bias the colour excess by ${\sim}0.05$ mag. This value is smaller than the typical colour excess uncertainty, suggesting that our analysis is robust to metallicity systematics of this scale (also see Section~\ref{sec:biases} and Fig.~\ref{fig:MH_error_M1}). Similarly, our analysis is insensitive to the stellar age of a star as long as it is on the MS. For instance, the colour index of an MS star less massive than ${\sim}1.1$ $\solarmass$ will change by less than $0.06$ mag as its age varies from 2 to 5 Gyr. 

However, inaccurate mass estimates may induce a more significant effect. Suppose the mass estimates have a systematic deviation of ${\sim}5$ per cent. Considering the mass range of primaries analysed in this work, such an error can induce a colour excess bias of up to around 0.25\,mag. The primary masses used in this study were taken from \textit{Gaia}'s \texttt{binary\_masses} table, described in \citet[][]{arenou22}. These estimates were calculated for \textit{Gaia}'s non-single-star catalogue, using the full orbital solution, the \textit{G}-band magnitude, and the \citet{Lallement_2019} extinction map. Their modelling procedure also assumed the companion is either non-luminous or a single MS star. The colour index, however, was not used in this process. 

The primary mass estimation procedure can sometimes result in overestimated primary masses accompanied by significant infrared colour excess. For example, in the case of a triple system, improperly modelled flux contribution from the astrometric secondary may increase the derived primary mass due to excess \textit{G}-band magnitude. This erroneous mass estimate may manifest as significant infrared colour excess. The expected deviations, on the order of ${\sim}0.25$ mag, are similar to the typical colour excess of the RCE sample (see Fig.~\ref{fig: color hist}). A possible fingerprint of the primary mass modelling inference in the RCE sample is discussed below.

\begin{figure}
        \centering
        \includegraphics[width=0.95\columnwidth]{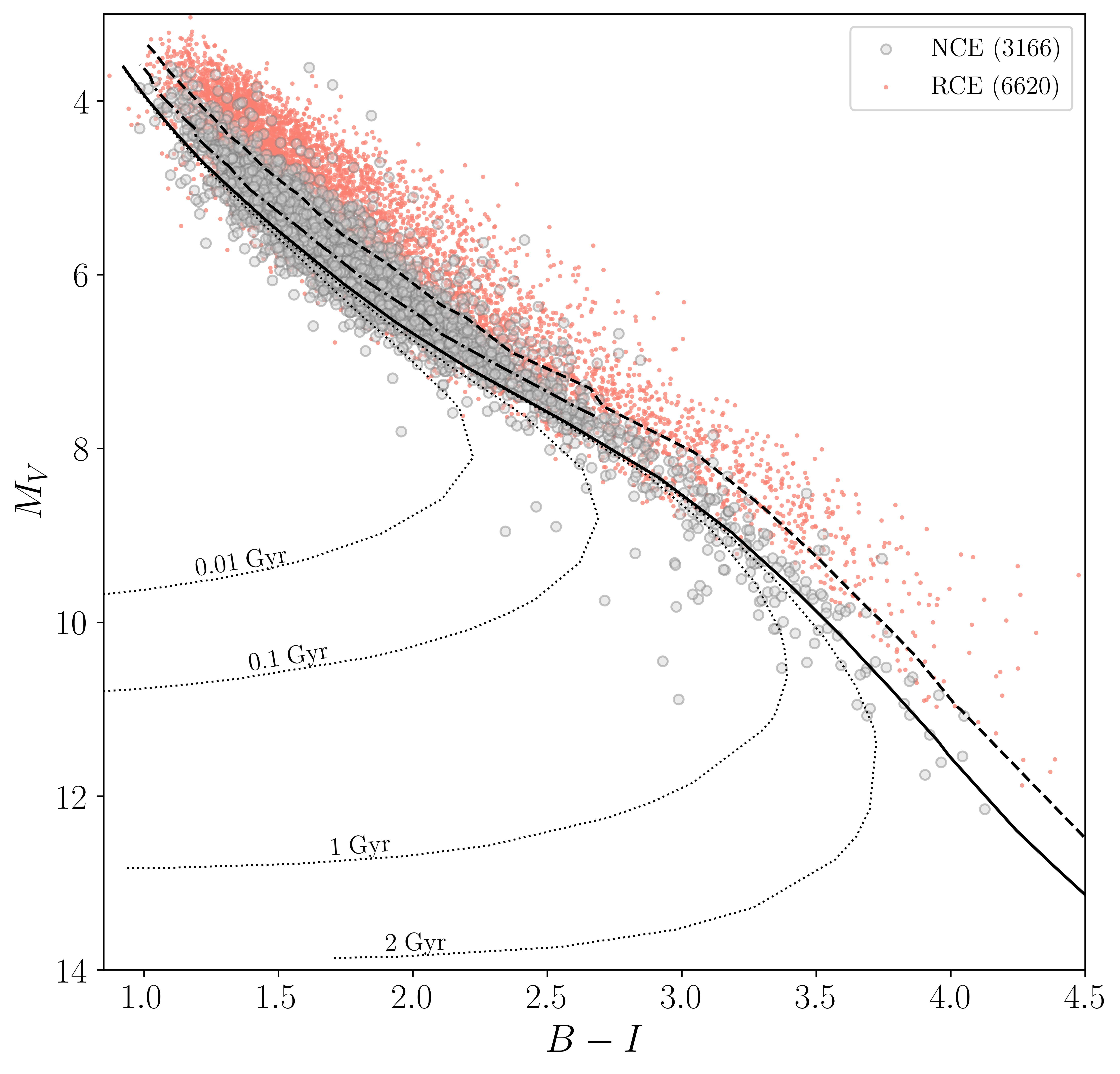}
        \caption{A colour-magnitude diagram for the $9\,786$ objects in our sample, not corrected for extinction.
        The red-colour excess (RCE) and no-colour excess (NCE) populations are plotted as red and grey points, respectively. The number of objects in each population is given in the figure legend. A 2~Gyr isochrone of $[{\rm M}/{\rm H}]=-0.2$ is plotted as a solid black line. The dashed and dash-dotted black lines illustrate the position of binary and hierarchical triple systems, respectively, and the dotted lines represent the position of MS+WD binaries at different WD cooling ages (see text).}
    \label{fig:CMD}
\end{figure}

\begin{figure*}
    \centering    \includegraphics[width=1\textwidth]{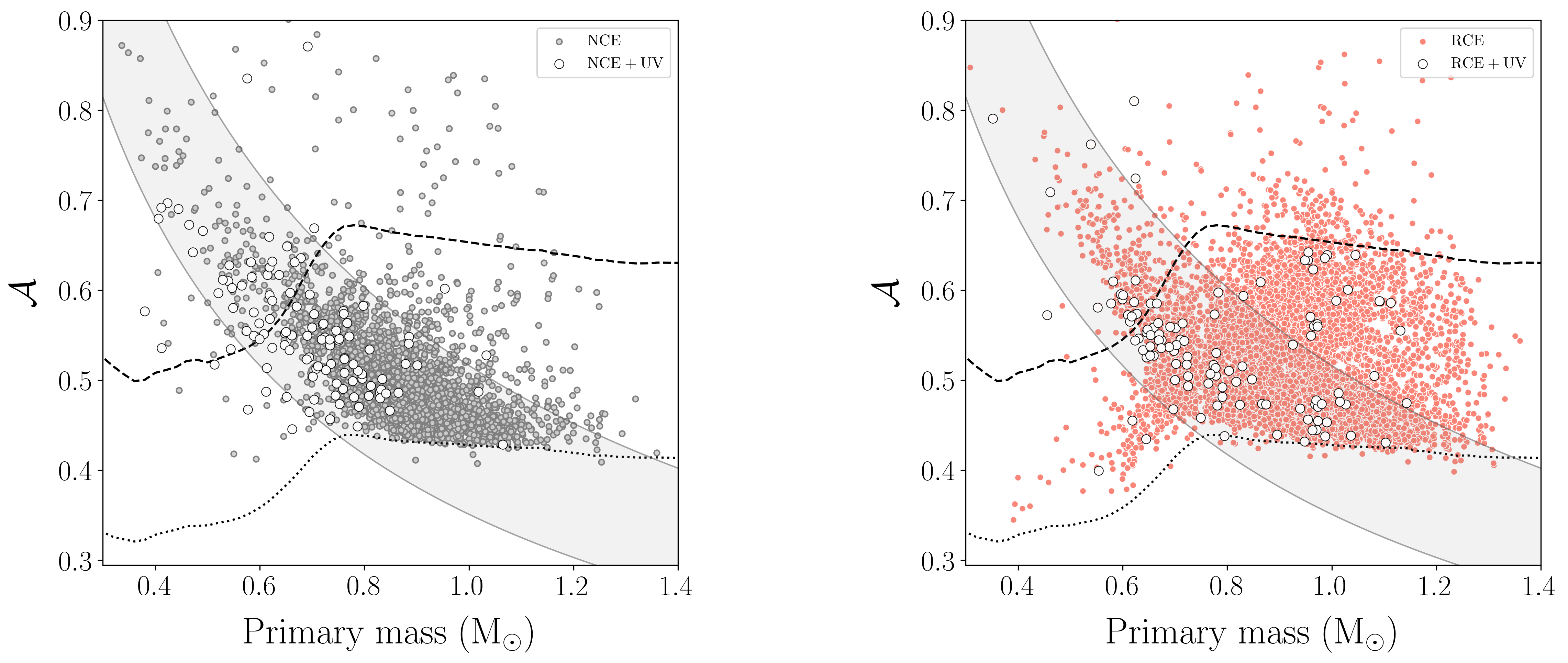}
    \caption{The AMRF versus the mass of the primary star, presented separately for the NCE (\textit{left} panel) and the RCE (\textit{right} panel) populations. The dotted line separates between \textit{class-I} and \textit{class-II} systems, and the dashed line separates \textit{class-II} from \textit{-III}. A grey strip highlights the expected location of systems with non-luminous companions in the mass range of $0.45{-}0.75\,\,\solarmass$, where typical MS+WD binaries are expected to reside. The figure shows that, as a population, NCE systems tend to populate this area in parameter space, as opposed to the RCE sample. The white points depict systems in which excess emission in the near-UV band of \textit{GALEX} was detected, as discussed in Section~\ref{sec: uv}, which we ascribe to the contribution of the WD companion. See Section~\ref{sec:biases} for a discussion of MS+WD contamination of the RCE sample.}
    \label{fig: AMRFvs m1}
\end{figure*}

\subsection{Partitioning by colour excess}
We define Pr(red) as the fraction of realisations where the dereddened observed colour index was larger than the expected colour excess. We have set the threshold above which a system is considered to have infrared excess to be the maximal Pr(red) value for which the measured colour excess is smaller than $10^{-2}$,
\begin{equation}
    {\rm Pr (red)} \geqslant 56~\text{per cent}.
\end{equation}
This threshold ensures that we are sensitive to the expected colour excess from all triple systems in our sample (see Fig.~\ref{fig: triple color excess}). We provide the colour-excess probability for all the targets in our sample in Table~\ref{tab: table1}.

Our non-\textit{class-I} sample includes 6\,620 systems with red colour excess (RCE). The remaining 3\,166 targets do not exceed the above-mentioned threshold and will be referred to as no-colour-excess (NCE) systems. Fig.~\ref{fig: color hist} shows the colour excess distribution of the entire population and the RCE and NCE subsamples.

Fig.~\ref{fig:CMD} presents our sample, with red points indicating RCE systems and grey points representing NCE systems. A black line shows a 2~Gyr isochrone with a metallicity of [M/H]~$=-0.2$, for demonstration (the colour-excess determination was performed individually for each system as described above). Using the same isochrone, we depict the positions of MS binaries, hierarchical triples, and MS+WD binaries. The black dashed line shows the location of binaries with a mass ratio of $0.8$; the dashed-dotted line represents triple systems with an equal-mass close binary as the faint astrometric companion, assuming a mass ratio of $0.8$ of the astrometric components; the dotted lines represent MS+WD binaries at various WD cooling ages, derived assuming a hydrogen-dominated WD with a mass of $0.6$~$\solarmass$ (using the synthetic colours of \citealt{Bergeron_1995, Holberg_2006, Kowalski_2006, Tremblay_2011, Blouin_2018, Bedard_2020}).\footnote{\label{fn: cooling track}\url{https://www.astro.umontreal.ca/~bergeron/CoolingModels/}}

The colour excess estimate was obtained using synthetic photometry in the JKC system. Other photometric systems, such as the SDSS system or \textit{Gaia}'s $G_\text{BP}$ and $G_\text{RP}$ bands, can also be used. We found that the JKC's $B-I$ colours are more sensitive to the colour excess than the broader \textit{Gaia} $G_\text{BP}$ and $G_\text{RP}$ bands, and are available for almost all targets in our sample, as opposed to the SDSS synthetic photometry in the $u$ band.

%=========================
\section{A Census of white dwarfs}
\label{sec: wds}
%=========================

In the previous section, we identified a sample of astrometric binaries where the companion is unlikely to be a single MS star. Within this sample, we used the $B-I$ colour index to identify targets exhibiting red colour excess, allowing us to identify the contribution of light from close binary companions in hierarchical triple systems. Thus, we now have a sample (referred to as NCE) in which the secondaries are unlikely to be 
\begin{enumerate}
    \item single MS stars, or
    \item close binaries of two MS stars.
\end{enumerate}
The following section shows that the NCE sample is consistent with a predominantly MS+WD population.

\subsection{Sample characteristics}
Fig.~\ref{fig: AMRFvs m1} shows the relationship between the AMRF and the mass of the primary star for the NCE and RCE populations. The dotted line separates between \textit{class-I} and \textit{-II} systems, and the dashed line separates \textit{class-II} from \textit{-III}. The grey stripes represent the expected position of non-luminous companions in the mass range of $0.45-0.75~\solarmass$. Notably, the two populations are differently distributed: The NCE sample is concentrated around the grey stripe, which corresponds to the expected mass range for WD companions; in contrast, the RCE sample is more spread out and populates a larger area on the $M_1-\mathcal{A}$ plane. 
The distribution of the NCE sample indicates that the population is mostly comprised of MS+WD binaries. However, the RCE sample is not a pure sample of hierarchical triples, as suggested by the RCE \textit{class-III} binaries found on the grey stripe. These are likely MS+WD binaries that were found to have an apparent colour excess (see below).

Fig.~\ref{fig:dAgb_hists} compares the distances, primary masses, $V$-band extinction, and Galactic latitudes of both populations. This figure shows that RCE systems are located at greater distances and closer to the Galactic plane than the NCE sample. The extinction histogram indicates that this may be due to our colour classification, as systems closer to the Galactic plane suffer from greater extinction and appear redder, suggesting that some of our extinction or metallicity estimates may be inaccurate. We further discuss this possibility in the following. An alternative explanation is that triple systems, which may be more luminous than MS+WD binaries, are observed at greater distances and show larger extinction values. The RCE primary-mass histogram in the top right panel of Fig.~\ref{fig:dAgb_hists} shows an excess of ${\sim}1\,\solarmass$ targets, accompanied by a deficiency of ${\sim}0.75\,\solarmass$ primaries (which is also apparent in the right panel Fig.~\ref{fig: AMRFvs m1}). These features do not appear in the NCE primary mass estimates. 

We speculate these artefacts stem from the primary mass estimation procedure, as they suggest that infrared excess is associated with its products. As mentioned above, the masses were derived using the absolute magnitudes of the targets without considering their colour index. Additionally, while the process did account for possible flux contribution from a single MS companion in a binary system, hierarchical triple systems were not modelled. It is, therefore, possible that triple systems were analysed, assuming their companion is non-luminous, in a process that affected their primary mass estimates, leaving a significant infrared colour excess signature.

To test whether the differences between the populations are caused by vulnerabilities in the orbital fitting scheme, we checked for systematic differences in the number of measurements (visibility periods used; VPU), the goodness of fit (GoF), and the significance of the angular semi-major axis ($\alpha_0 /\Delta \alpha_0$) and parallax ($\varpi /\Delta \varpi$). The corresponding histograms are plotted in Fig.~\ref{fig:fit_info_hists}. While the two populations follow similar trends in these four parameters, the NCE population seems to have slightly better significance and GoF values. This is expected, considering the larger distances and lower Galactic latitudes of the RCE sample.

\begin{figure*}
    \begin{minipage}{0.475\textwidth}
        \centering
        \includegraphics[width=0.9\columnwidth]{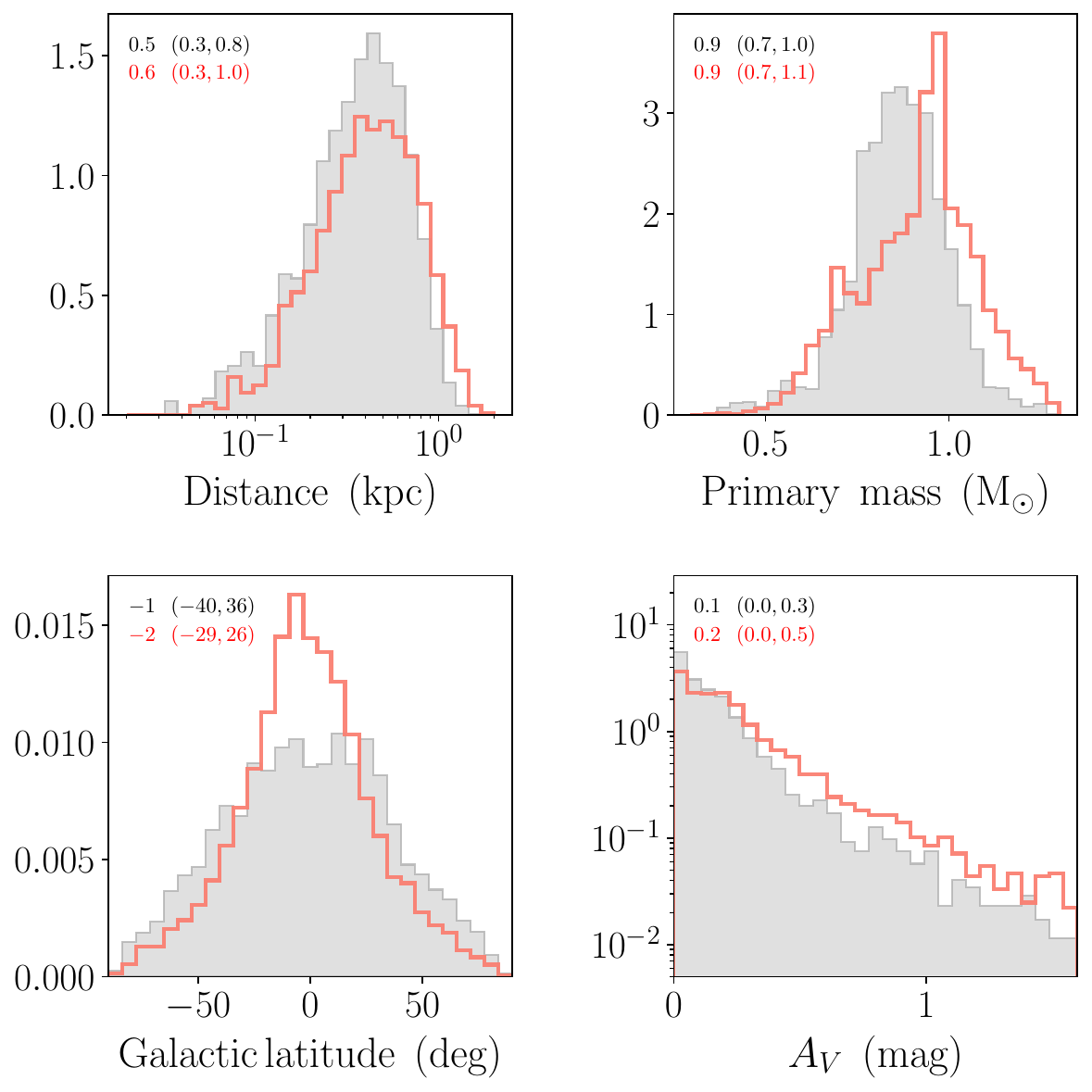}
        \caption{Distance, primary mass, $V$-band extinction, and Galactic latitude histograms (\textit{clockwise from top left}) for the RCE sample (red line) and the NCE sample (grey bars). The median and the $36-84$~percentile range for each distribution are written on the top left corner of each panel, where the upper and lower rows describe the NCE and RCE samples, respectively.
        }
        \label{fig:dAgb_hists}
    \end{minipage} \hfill
    \begin{minipage}{0.475\textwidth}
        \centering    
        \includegraphics[width=0.9\columnwidth]{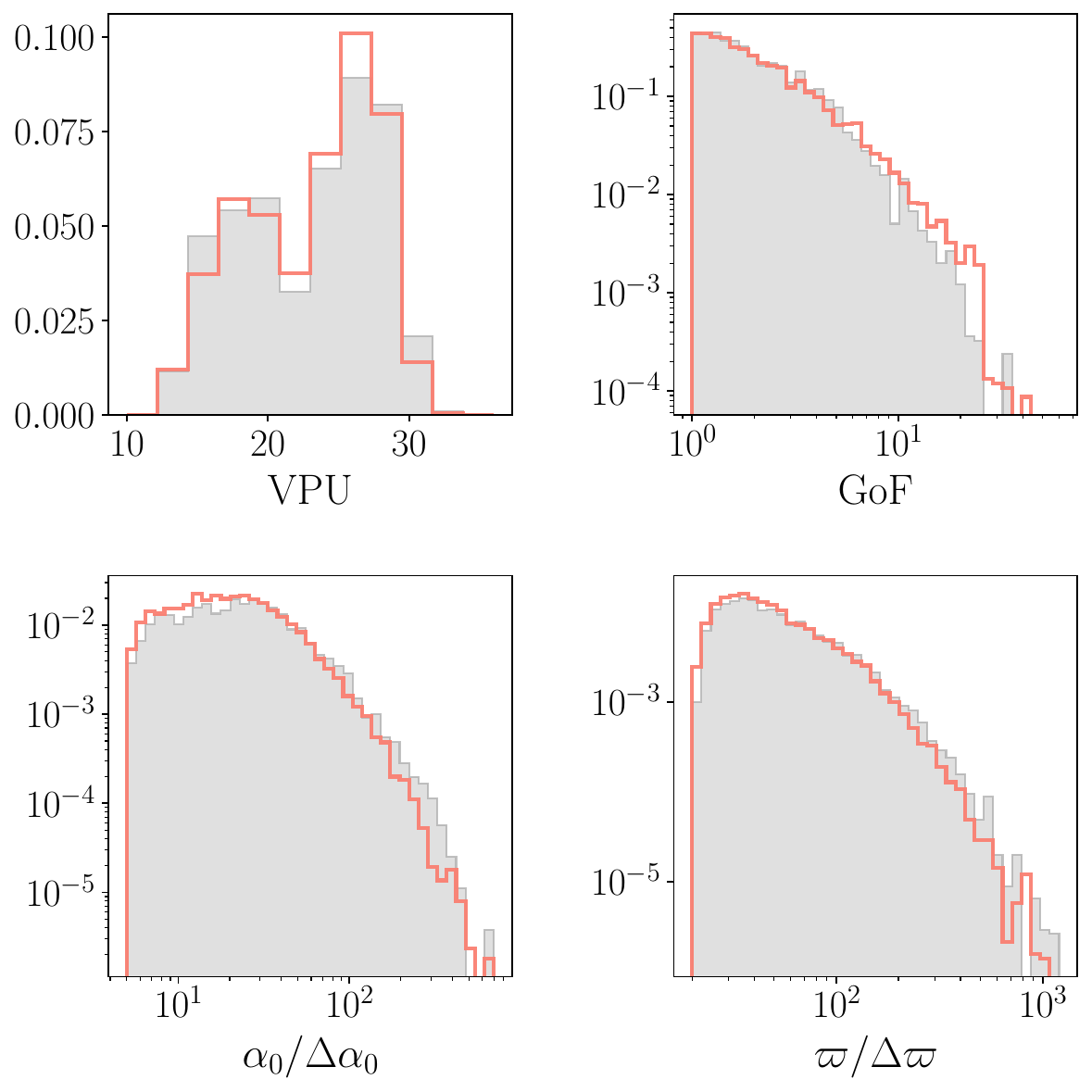}
        \caption{Properties of the selected sample. The two \textit{top} panels show the number of visibility periods used (VPU; \textit{left}) and the goodness of fit (GoF; \textit{right}) reported in the \textit{Gaia} catalogue. The significance of the angular semi-major axis (\textit{left}) and parallax (\textit{right}) are shown in the \textit{bottom} panels. The colour coding is the same as in Fig.~\ref{fig:dAgb_hists}.  }
        \label{fig:fit_info_hists}
    \end{minipage}
\end{figure*}
 
\begin{figure*}
    \centering    \includegraphics[width=1\textwidth]{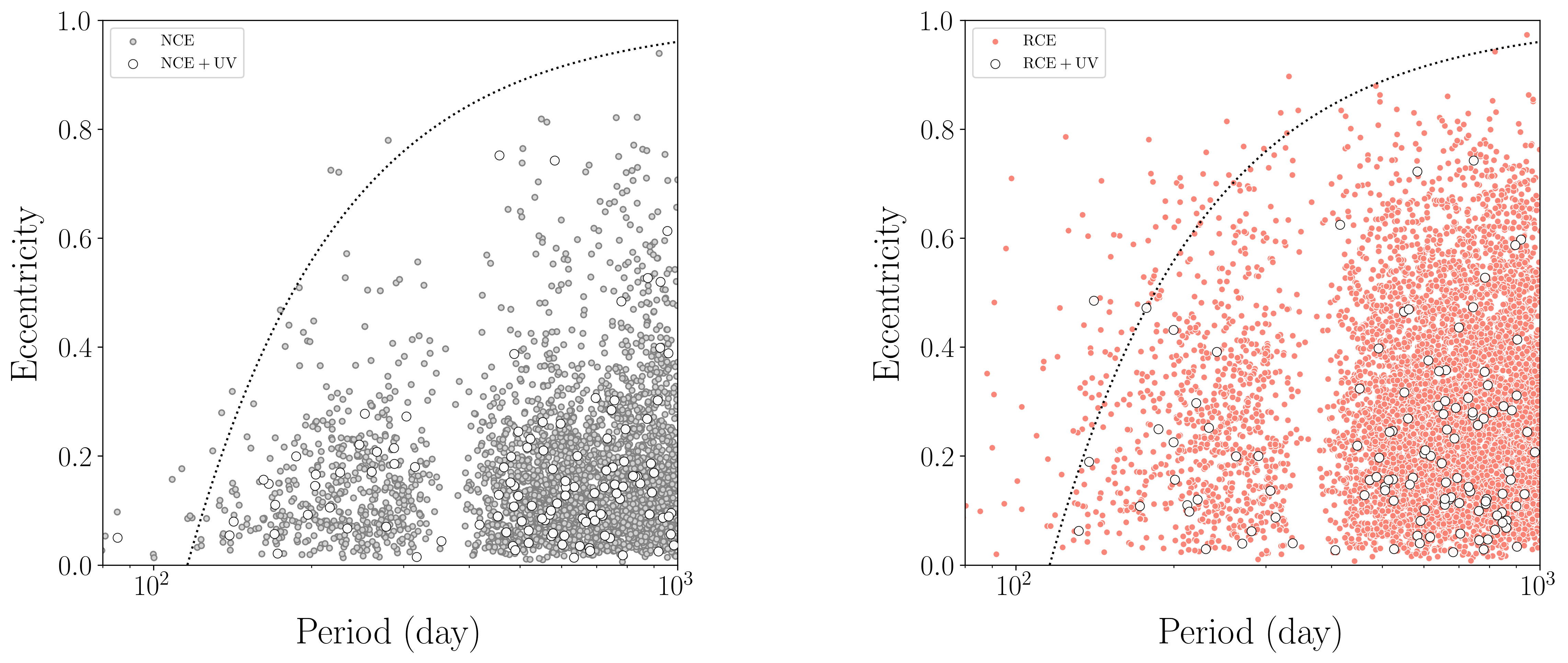}
    \caption{A period-eccentricity diagram for the NCE and RCE populations is plotted on the left and right panels, respectively. The gap corresponds to orbital periods of about one year. The dotted line represents an upper envelope that corresponds to periastron distance of ${\sim}100~{\rm R}_\odot$, assuming a total mass of $1~\solarmass$. The white points depict systems in which excess emission was detected (see Figure~\ref{fig: AMRFvs m1} and Section~\ref{sec: uv}). }
    \label{fig: P-e diagram}
\end{figure*}

\begin{figure}
       \centering\includegraphics[width=0.825\columnwidth]{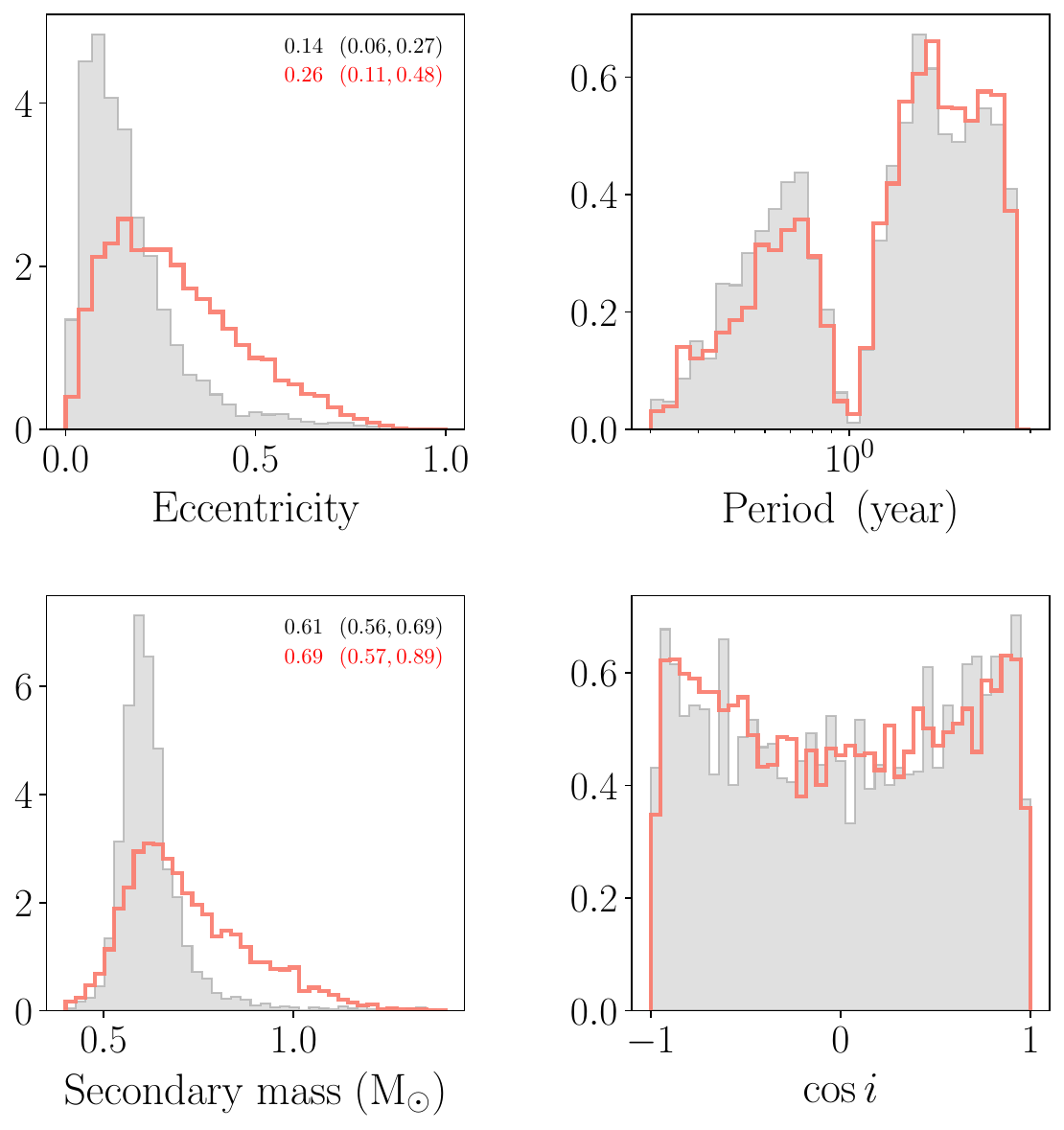}
        \caption{Histograms of the eccentricity, orbital period, cosine of the inclination angle, and secondary mass (\textit{clockwise from top left}). The colour coding is the same as in Fig.~\ref{fig:dAgb_hists}. The two left panels, showing the eccentricity and secondary mass, also show the median and $36-84$ percentile range for the NCE (top line) and RCE (bottom line) populations.} \label{fig:orbital_params_hists}
\end{figure}

Fig.~\ref{fig: P-e diagram} presents the eccentricity versus the orbital period of the stars in the sample. 
Most of the eccentricities of the RCE sample (shown in the right panel)  are below ${\sim}0.8$. 
This is probably an observational bias caused by the significant amount of time highly eccentric systems spend close to apastron, where the motion is not easily detected.
Furthermore, most of the RCE binaries are found below the dotted line that indicates orbits that reach separations of ${\sim}100\, {\rm R}_\odot$ at periastron passage, assuming a total mass of $1\,\solarmass$. On the other hand, the eccentricities of the NCE sample (shown in the left panel) are mostly below ${\sim}0.3$, as can be seen in the left panels of Fig.~\ref{fig:orbital_params_hists}.

The secondary mass distribution of the NCE targets,  shown in Fig.~\ref{fig:orbital_params_hists}, peaks sharply around ${\sim}0.6 \, \solarmass$, consistent with them being MS+WD binaries. In contrast, the RCE sample displays a broader secondary mass distribution peaking at ${\sim}0.7 \, \solarmass$. However, even if we assume that the RCE sample is purely comprised of triple systems, inferring the secondary mass or mass-ratio distribution of hierarchical triple systems based on it is not straightforward. The secondary mass estimates assume negligible light contribution from the astrometric secondary (see table~1 in \citetalias{shahaf23}). This assumption is invalid for hierarchical triple systems, making these estimates merely lower limits of the secondary mass \citep{shahaf19}.

Fig.~\ref{fig:orbital_params_hists} also presents histograms of the orbital periods and the cosine of the inclination angle. Compared to the NCE population, the occurrence of systems with orbital periods below one year is smaller in the RCE population. One simple explanation is that RCE systems are found at greater distances, where relatively close separations are harder to detect. However, this difference may also reflect the interaction between the present-day primary MS and the WD progenitor when it was a red giant or, alternatively, the stability of hierarchical triple systems. In terms of the inclination angle, the two populations exhibit similar distributions.

%---------------------------
\subsection{UV excess}
\label{sec: uv}
%---------------------------

Newly formed WDs can significantly contribute to the binary system's brightness in the visible bands (e.g., Fig.~\ref{fig:CMD}). This contribution becomes more pronounced in the UV, where MS primaries below ${\sim}1.2$~$\solarmass$ are relatively faint. To identify these WD contributions, we searched the Galaxy Evolution Explorer (\textit{GALEX}; \citealt{morrissey07}) database for UV counterparts to the NCE binaries in our sample. A similar analysis of the \textit{class-III} binaries from \citetalias{shahaf23} was recently done by \citet{Ganguly2023}.

We cross-matched the NCE targets in our sample with the \textit{GALEX} all-sky imaging survey (AIS; 
\citealt{bianchi11, bianchi17}). The \textit{Gaia} coordinates were propagated back to the coordinates at the AIS midpoint using the measured proper motion of each system. The fiducial search cone radius around each coordinate was $3$\,arcsec \citep{Bianchi_2020}. This cone was enlarged to account for the proper motion of each system during the \textit{GALEX} mission duration. 

We have identified $3\,673$ matches between \textit{GALEX} and \textit{Gaia} targets, out of which $1\,436$ belong to the NCE population. Table~\ref{tab: table2} provides a list of the matched sources, including the \textit{Gaia} and \textit{GALEX} identifiers, apparent UV magnitude, and the angular separation between the matched sources. The median separation between the two positions of the matched sources is ${\sim}0.6$\,arcsec. Only $7$ sources are separated by $3$ arcsec or more, with a maximal separation of ${\sim}3.6$\,arcsec. There are $9$ targets for which we identified a second source within the search cone; we chose the closest one in these cases. 

We use a simple single-valued relation to estimate the extinction in the \textit{GALEX} near-UV (NUV) band.\footnote{\textit{GALEX NUV}, spanning the range of $1771{-}2831\,{\angstrom}$.} The relation is given by
\begin{equation}
A_{\rm NUV} \approx \frac{R_{\rm NUV}}{R_{\rm G}} A_{\rm G} \, \simeq 3.085 \, A_{\rm G},
\end{equation}
where $A_{\rm NUV}$ and $A_{\rm G}$ represent the extinction in the NUV and \textit{G} bands, respectively. Their corresponding extinction coefficients, $R_{\rm NUV}$ and $R_{\rm G}$, were obtained from \citet{zhang23b}. 

\begin{figure}
    \centering    \includegraphics[width=0.9\columnwidth, trim={0.25cm 0 0 0 },clip]{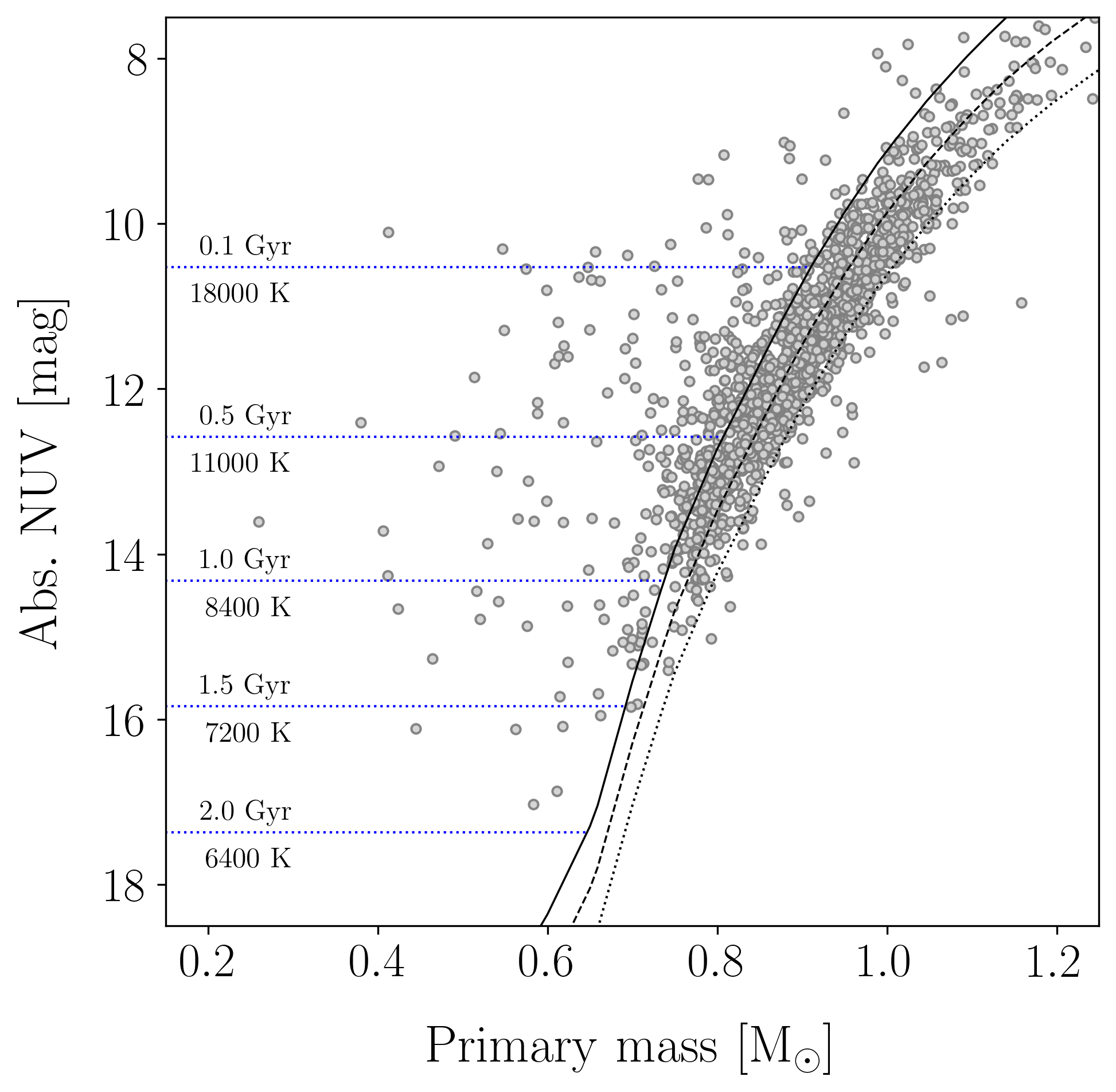}
    \caption{The GALEX NUV absolute magnitude versus the primary star's mass. The grey circles represent the NCE binaries we have identified near-UV counterparts in \textit{GALEX} (see text). The solid, dashed and dotted black lines represent the expected MS relation, based on a PARSEC isochrone of 2\,Gyr, for extinction values of $0$, $0.75$ and $1.5$ mag, respectively. The horizontal lines show the expected NUV absolute magnitude of a $0.6\,\solarmass$ WD at various cooling ages.}
    \label{fig: nuv hrd}
\end{figure}

Fig.~\ref{fig: nuv hrd} presents the NUV absolute magnitudes of the NCE sample. The solid black line depicts the corresponding MS mass-luminosity relation obtained using our default isochrone (see above). The PARSEC NUV magnitudes are provided in the Vega photometric system; therefore, the derived relation was shifted by 1.699\,mag to match the AB system of the AIS (\citealt{bianchi11b}). To illustrate the effect of extinction on the UV emission from the system, we added dashed and dotted black lines representing an extinction of $0.75$ and $1.5$\,mag, respectively. These values roughly correspond to the 80\textsuperscript{th} and 95\textsuperscript{th} percentiles of the NCE sample.

Most points in Fig.~\ref{fig: nuv hrd} fall between the solid and dotted black lines, as expected. To illustrate the expected contribution of the WD, we plotted as dotted blue lines the expected absolute NUV magnitudes of $0.6~\solarmass$ WDs with hydrogen-dominated photospheres at different cooling ages. For this purpose we used the tables\footref{fn: cooling track} of WD synthetic colours by \citet{Bergeron_1995, Holberg_2006, Kowalski_2006, Tremblay_2011, Blouin_2018}; and \citet{Bedard_2020}. The diagram suggests that WDs with cooling ages below ${\sim}0.5$ Gyr should be easily detectable if their primary MS companions are of spectral types later than K.

Around 8~per cent of the NCE sample, 117 targets, show significant excess UV emission. As with the infrared colour excesses, we somewhat arbitrarily defined excess UV as a system brighter than expected MS single-star curve by more than $1\sigma$ (see Table~\ref{tab: table2}). We accounted for the NUV magnitude, parallax, extinction and primary mass uncertainties. The RCE sample, for comparison, contains 118 binaries (${\sim}5$~per cent) showing NUV excess detected by the same criterion. However, nearly $40$ of these systems have primaries with masses of ${\sim}1$ \solarmass, which is unlikely, given the young WD cooling ages implied. Considering the RCE primary mass distribution in Figure~\ref{fig:dAgb_hists}, it is likely that the expected NUV value of some of these systems is erroneous. The fraction of RCE systems with NUV excess is probably closer to ${\sim}3.5$~per cent.

As discussed in the following section, the colour-based sample segmentation aims to ensure that the NCE sample is less contaminated by triple systems, not vice versa. Therefore, we expect the RCE sample to contain triple systems and MS+WD binaries.
Fig.~\ref{fig: AMRFvs m1} shows that the NUV-bright systems are localised around the WD grey stripe on the $\mathcal{A}{-}M_1$ diagram, both for the NCE and RCE samples, as expected.
The higher occurrence of NUV-bright binaries in the NCE sample and the consistency of their secondary masses with the typical mass of a WD corroborates our classification scheme. The ratio between the two samples suggests that the contamination of the RCE sample can reach, at worst, about $60$~per cent. Still, a more detailed analysis of the spectral energy distribution of each system is required \citep[e.g.,][]{Ganguly2023}.

\section{Biases and selection effects}
\label{sec:biases}

\begin{figure}
    \centering
    \includegraphics[width=\columnwidth]{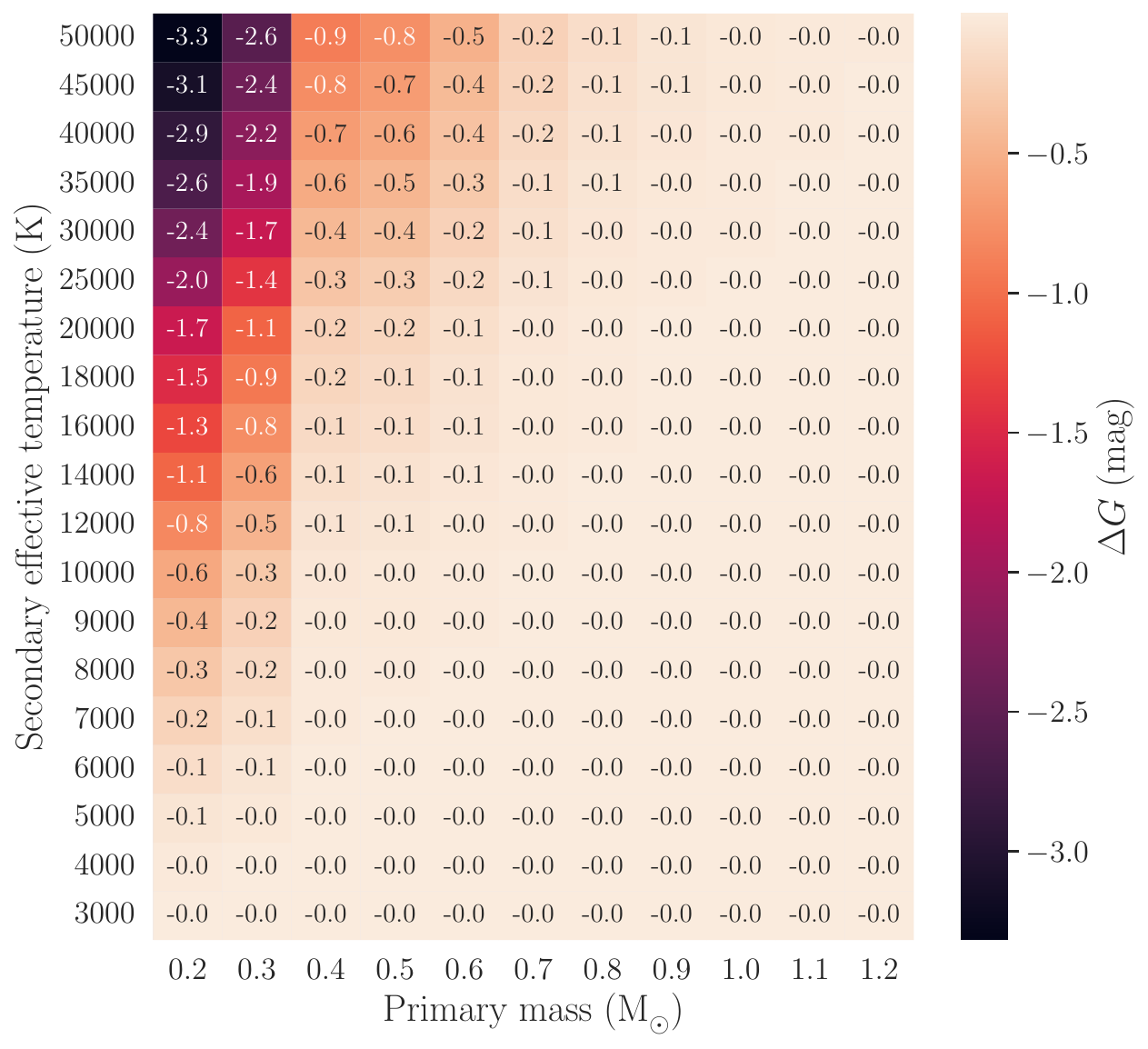}
    \caption{The expected photometric excess in the \textit{Gaia} $G$ band of an MS+WD binary compared to the expected emission from the primary MS alone. This calculation assumes black-body spectra for both components and a 0.6\,\solarmass\ hydrogen-dominated DA WD.}
    \label{fig:PhotometricExcess_G}
\end{figure}

The first obvious selection effect in our NCE sample stems from the pre-selection of primary MS stars since low-mass MS stars will appear as photometric primaries in the \textit{Gaia} $G$ band only if their WD companion is cool enough. However, this will affect only the smallest MS primaries, as demonstrated in Fig.~\ref{fig:PhotometricExcess_G}, which shows the expected photometric excess in the \textit{Gaia} $G$ band of an MS+WD binary compared to the expected emission from the primary MS alone, as a function of primary mass and WD effective temperature. From this figure it is clear that only binaries with a $\sim 0.3$\,\solarmass\ MS star and a WD warmer than $\sim 16\,000$\,K, or a $\sim 0.2$\,\solarmass\ MS star and a WD warmer than $\sim 12\,000$\,K, would be a priori excluded from our sample, since the WD would dominate the \textit{Gaia} $G$ band. This explains the sparsity of systems with smaller primary masses (see Fig.~\ref{fig: AMRFvs m1}). The expected photometric excess was calculated assuming black-body spectra for both components. The radius of the MS component was estimated using a 2\,Gyr, [M/H]~$=0$, PARSEC isochrone, while the WD radius was estimated using the \texttt{WD\_models}\footnote{\url{https://github.com/SihaoCheng/WD_models}} \texttt{Python} package with the models of \citet{Bedard_2020}, assuming a 0.6\,\solarmass\ hydrogen-dominated DA WD.
 
\begin{figure}
        \centering\includegraphics[width=0.9\columnwidth]{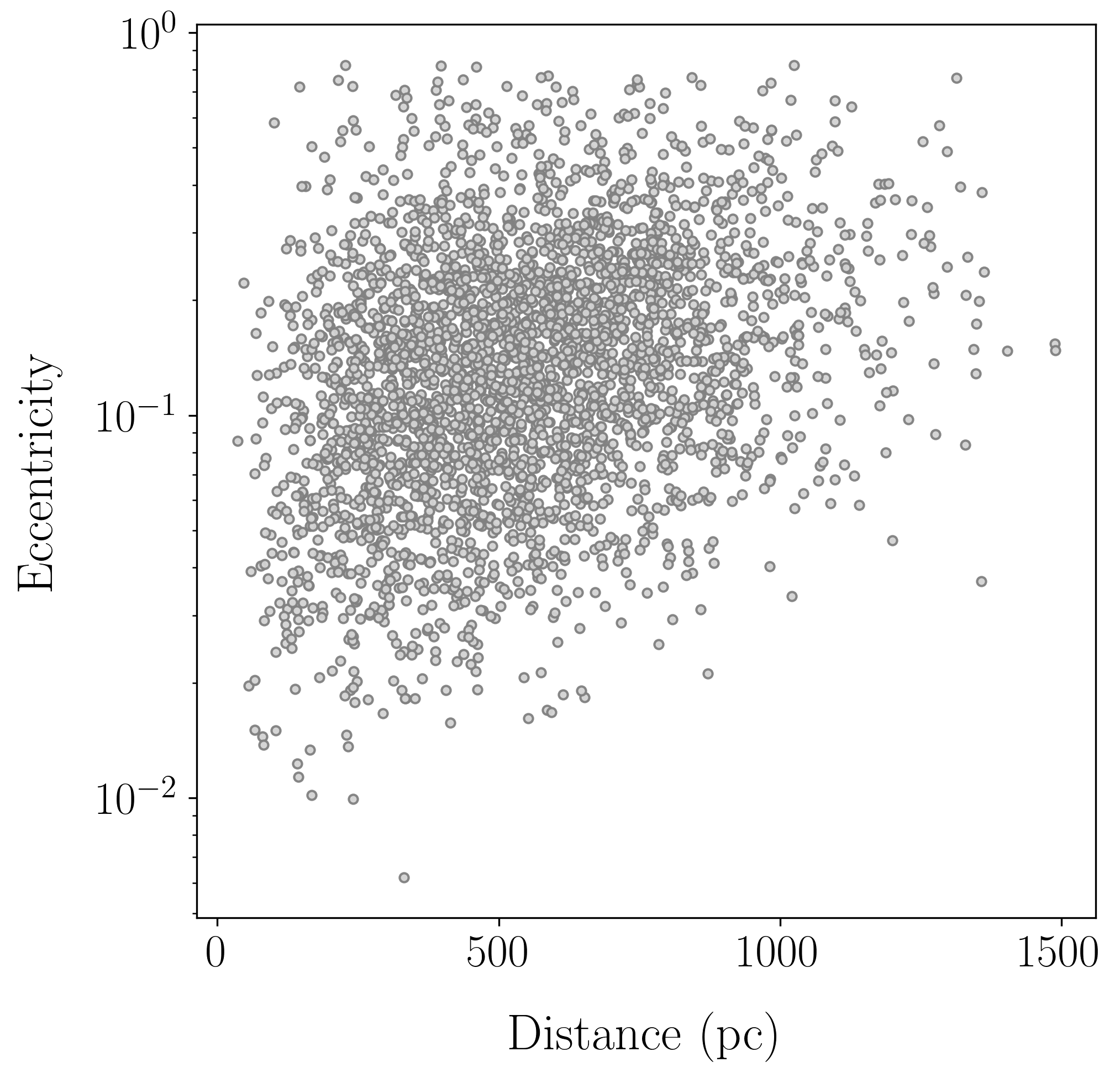}
        \caption{ The orbital eccentricity versus the parallax distance for the systems in the NCE sample.}
        \label{fig: dist ecc}
\end{figure}

Our NCE sample, presumably primarily consisting of systems with WD companions, suggests that many MS+WD binaries exhibit a modest eccentricity of ${\sim}0.15$. This eccentricity might be linked to processes during the final stages of stellar or binary evolution (see Section~\ref{sec:BinaryEvolution} below). Thus, it is noteworthy that the eccentricity distribution of our sample is significantly biased. Fig.~\ref{fig: dist ecc} presents the orbital eccentricity versus the system's distance, indicating a tendency towards higher eccentricities with increasing distances. One explanation for this bias could be that the eccentricity is derived more accurately at closer distances; if the inference procedure is sensitive to correlated noise, this situation can result in biased eccentricity estimation (\citealt{lucy71, bashi22}; also see, for example, a similar discussion by \citealt{hara19} for the case of radial-velocity modelling of exoplanetary orbits). However, a comprehensive analysis to determine the origin of this bias requires a detailed study of \textit{Gaia}'s selection function, which falls outside the scope of this study.

\begin{figure}
        \centering\includegraphics[width=0.9\columnwidth]{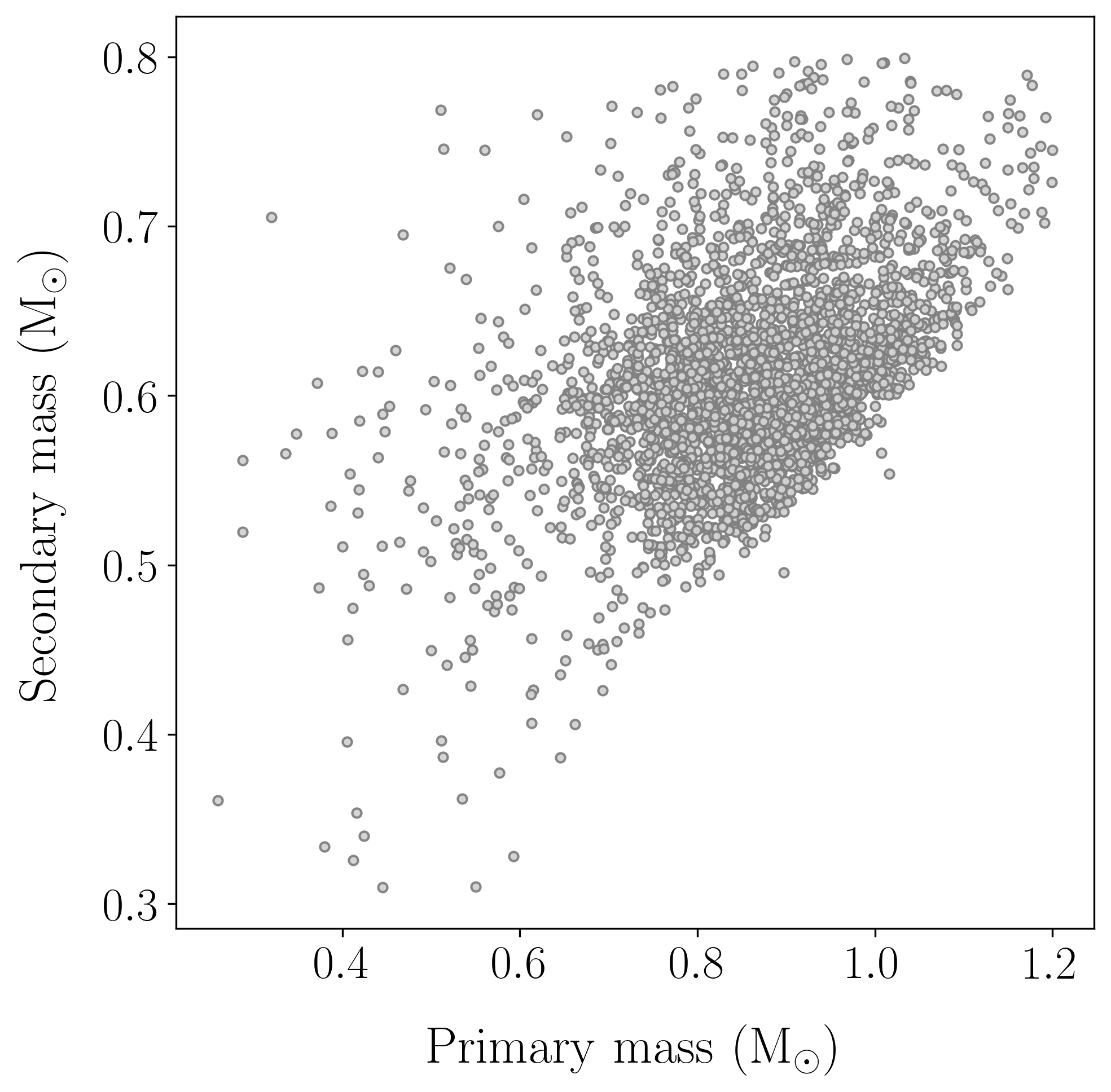}
        \caption{The secondary versus primary masses of the systems in the NCE sample. Systems with secondary masses larger than $0.8$~$\solarmass$ are not included in this diagram.}
        \label{fig: m1 m2}
\end{figure}

Another selection bias is related to the masses of the primary stars. The triage classification limits vary as a function of the primary star's mass (see Fig.~\ref{fig: AMRFvs m1}, for example). As a result, the WD mass distribution is biased since the minimal WD mass grows with the mass of the primary star, as Fig.~\ref{fig: m1 m2} illustrates. One immediate consequence of this selection effect is a bias towards high WD masses and high mass ratios in the NCE sample. However, other biases may also stem from this selection effect. For example, since all the primary stars in our sample are on the MS, the correlation between the secondary and primary masses can be translated into a correlation between the WD's mass and the binary system's absolute magnitude. As a result, the maximal distance---and hence the orbital separation and eccentricity---of the detected systems might also be affected.

\begin{figure}
    \includegraphics[width=0.95\columnwidth, trim={0 0 0.25cm 0},clip]{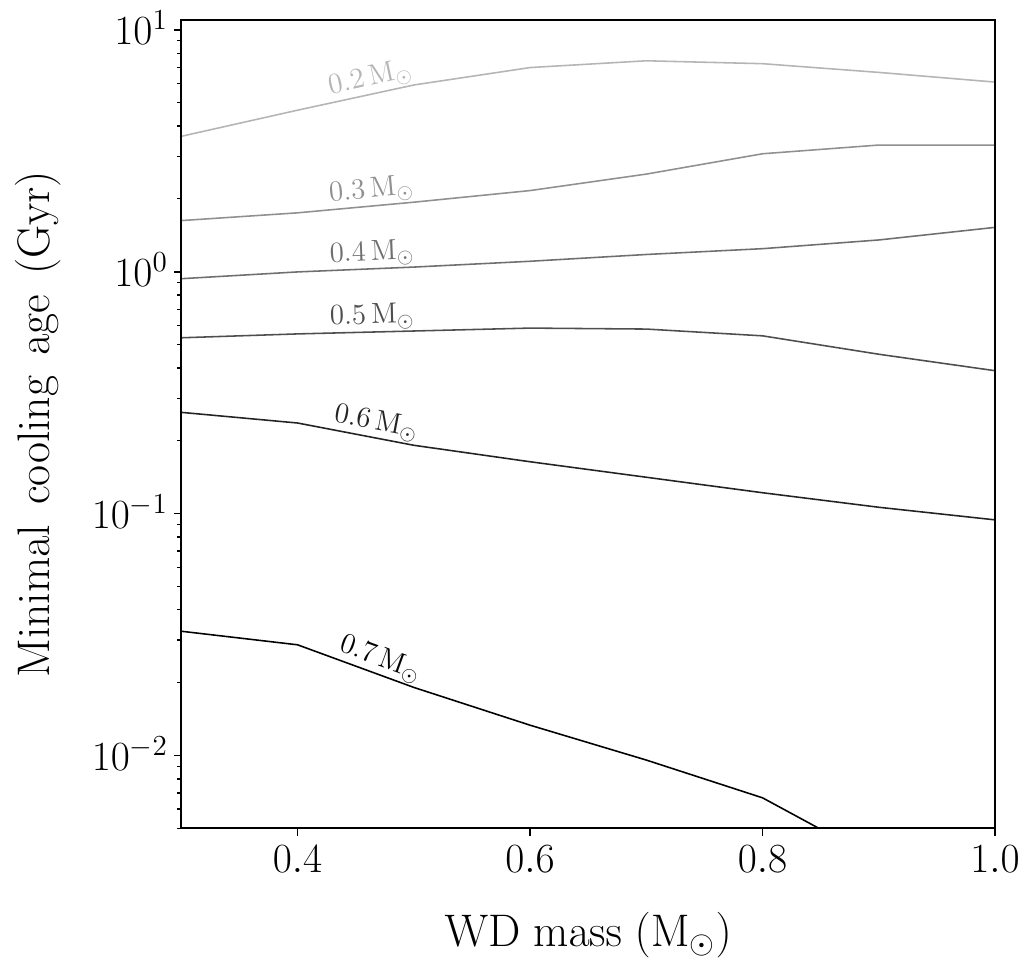}
        \caption{The minimal WD cooling age for which the $G$-band flux ratio between the WD and the MS drops below $5\%$ as a function of the WD mass. Each line represents a system with a different primary MS mass.}
        \label{fig: wd_mass_cooling_age}
\end{figure}

Finally, the assumption that the light contribution from the WD is negligible compared to that of the MS primary star is not necessarily accurate. Despite their compact nature, their contribution to the overall luminosity of the binary system may be non-negligible. If unaccounted for, this contribution may bias the WD mass estimates. The relative intensity of the WD compared to its MS companion depends on several factors, including the mass of the MS primary, the mass of the WD itself, and its cooling age. Since WDs cool with time over billions of years, younger WDs are brighter and bluer. In Fig.~\ref{fig: wd_mass_cooling_age}, we plot the minimal cooling age below which WDs contribute more than $5$~per cent of the system's light in \textit{Gaia}'s bandpass. WDs younger than this critical age can introduce bias more significant than ${\sim}10$~per cent to the mass estimation process. For systems with $\gtrsim 0.5$\,\solarmass\ MS primaries and WDs with cooling ages older than ${\sim}0.5$ Gyr, the contribution of light from the WD cannot distort their mass estimates significantly.


\begin{figure}
    \centering
    \includegraphics[width=0.9\columnwidth]{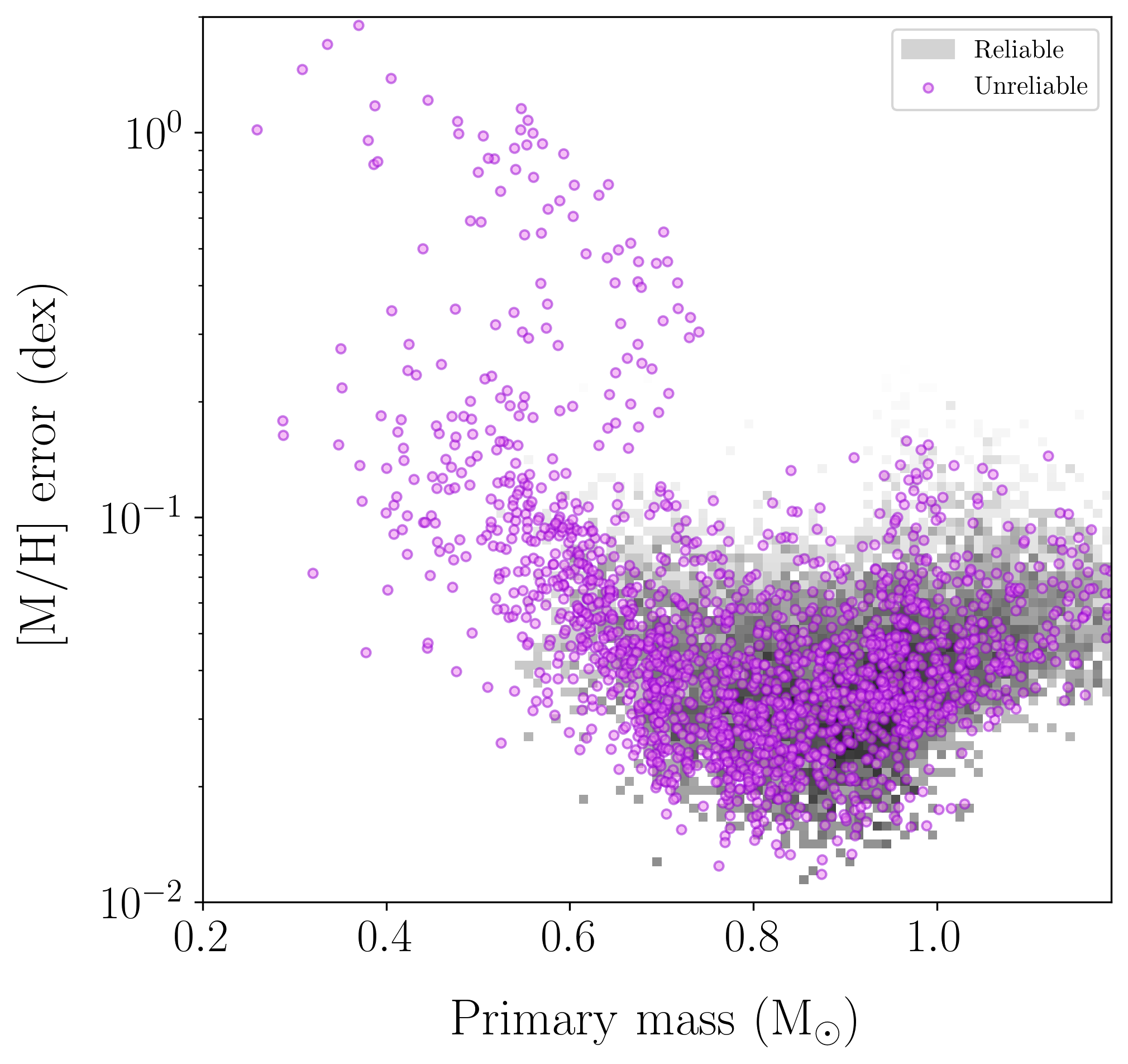}
    \caption{The metallicity error as a function of primary mass of all non-\textit{class-I} systems. Reliable [M/H] estimates ($\texttt{quality\_flags} < 8$ in \citealt{Zhang_2023}) are shown in the greyscale two-dimensional histogram. Unreliable [M/H] estimates ($\texttt{quality\_flags} \geq 8$) are plotted as violet circles. These systems were assigned a metallicity of $0\pm0.25$\,dex in our analysis (see Section~\ref{sec: colour excess}).}
    \label{fig:MH_error_M1}
\end{figure}

It is important to mention that the RCE sample does include some MS+WD binaries. For example, some of the RCE binaries are located above the \textit{class-III} limit and within the region of the WD mass stripe of Fig.~\ref{fig: AMRFvs m1}. Another indication of the WD contamination in the RCE sample is the NUV-bright systems. The position of these systems on the diagram is generally consistent with the WD stripe. Hence, these are likely misclassified MS+WD binaries identified as having excess infrared emission. This misclassification is probably the result of erroneous metallicity estimates for systems with primary masses smaller than $\sim0.6$\,\solarmass\ (see Fig.~\ref{fig:MH_error_M1}) that lead to an unreliable colour-excess assignment. This is consistent with fig.~18 of \citet{Zhang_2023}, demonstrating the low fraction of reliable stellar parameter estimates for M dwarfs that were not included in the LAMOST training set used in that study.
Luckily, almost the entire expected range of M-dwarf primaries with WD secondaries resides within \textit{class III} on the AMRF plot (Fig.~\ref{fig: AMRFvs m1}). Thus, all \textit{class-III} systems with primary masses $\lesssim 0.6$\,\solarmass\ are very likely to be MS+WD binaries, regardless of their assigned colour excess. Since there are very few \textit{class-II} systems with M-dwarf primaries, the impact of this issue on our final MS+WD catalogue is small and can affect only systems with extremely low-mass WD companions. Better estimates of M-dwarf metallicities will enable more accurate classifications of these systems in the future.

\textit{Class-III} systems with primaries more massive than $\sim0.6$\,\solarmass\ that are included in the RCE sample are more likely to have erroneous primary mass estimates (especially those with masses around $\sim 1$\,\solarmass, see discussion above). Alternatively, the assumption of 2\,Gyr-old isochrones used for the colour-excess estimation (see Section~\ref{sec: colour excess}) might be inaccurate for the systems with the most massive primaries, which could be either younger or already in the process of leaving the MS. Finally, we mention that the RCE sample tends towards lower Galactic latitudes. In more crowded fields, excess flux from foreground or background sources could potentially impact the mass estimates, orbital fitting, or colour excess measurements. It is possible that crowding contributed to the artefacts detected in the primary mass histogram and affected the GoF distributions of the orbital solutions (see Fig.~\ref{fig:dAgb_hists} and~\ref{fig:fit_info_hists}, respectively).

%=========================
\section{Summary and outlook}
\label{sec: summary}
%=========================

\subsection{Astrophysical implications}
The astrometric binaries presented here provide a uniquely large sample of nearly $3\,200$ probable MS+WD binaries with orbital separations of ${\sim}1$~au. These systems populate a region of parameter space largely unexplored by other observational techniques---only a few MS+WD binary systems with orbital periods of hundreds of days are currently known (see \citealt{Anguiano_2022, Parsons_2023}, and references therein), enlarging the sample size by about two orders of magnitude. This sample has already yielded some insights into the WD population and binary evolution \citep{Hallakoun_2023}.

\subsubsection{Binary evolution}\label{sec:BinaryEvolution}

On the one hand, the present orbits of many of the binaries are too small to allow for uninterrupted evolution of the WD progenitors through their red-giant phase; 
the eccentricity-period diagram displays some binaries with large eccentricities, such that the binary separation at the periastron is smaller than, say, $100~R_{\odot}$. 
On the other hand, their separation is likely too large to assume that they went through a common-envelope phase, as post-common envelope binaries with a WD and M-dwarf components have been found to exist only at the shortest periods (less than about two weeks, peaking at ${\sim}8$\,hr; \citet{2011A&A...536A..43N, 2019MNRAS.484.5362A, 2021ApJ...920...86K, 2022MNRAS.512.2625L, Roulston_2021}, but see also \citet{Yamaguchi_2024} for some systems with periods of up to seven weeks). 
Therefore, some of these binaries, after the WD nature of the unseen companion is confirmed, might necessitate special binary evolutionary tracks \citep[e.g.][]{2009ApJ...697.1048P}.

Stable mass transfer is a known mechanism to form long-period binaries with WD components. In fact, theory predicts a relation between the period and the WD mass for systems that have evolved this way \citep{1995MNRAS.273..731R,Chen_2013}. For WD masses between $0.45-0.5\,\solarmass$ and donor stars on the Red Giant Branch (RGB), \cite{Chen_2013} predicts periods between $600-935$~days and $275-428$~days at metallicities of $z=0.01$ and $z=0.0001$, which are in good agreement with the periods of our sample. However, most of our WD masses are above $0.55\,\solarmass$, so their direct progenitors cannot be RGB stars but must have already been on the Asymptotic Giant Branch (AGB) at the onset of the mass transfer. While mass transfer with an AGB donor is typically expected to lead to a common-envelope phase \citep[e.g.][but see \citealt{2007ASPC..372..397M}]{1987ApJ...318..794H, 2002MNRAS.329..897H}, our systems imply stable mass transfer is possible, likely aided by mass loss through non-conservative mass transfer \citep{1997A&A...327..620S}. 
In addition, the eccentricities of our systems cannot easily be explained by stable mass transfer alone since, in classical binary evolution theory, tides would circularise the orbit before the onset of the mass transfer. Possible eccentricity pumping mechanisms are enhanced wind mass loss at periastron \citep{1995A&A...293L..25V,2008A&A...480..797B}, Roche-lobe overflow at periastron \citep{2000A&A...357..557S,2015A&A...579A..49V}, formation of a circumbinary disc \citep{1996A&A...315L.245W, 2013A&A...551A..50D,2015A&A...579A..49V}, and WD natal kicks \citep{2010A&A...523A..10I}. 

If some of these binaries underwent mild mass-transfer interaction, we might find evidence for heavy s-process element pollution in the luminous MS star atmosphere. Very few known barium stars are in our sample, and more generally in the \textit{Gaia} non-single star catalogue. Interestingly, the period-eccentricity distribution in the left panel of Fig.~\ref{fig: P-e diagram} resembles that seen in barium stars at a similar range of orbital periods (e.g., \citealt{2016A&A...586A.158J, 2019A&A...626A.127J, escorza23}). The orbital elements and amount of enrichment may uncover the properties of the WD progenitors and the underlying mass-transfer process. 
\citet{sayeed23}, for example, recently proposed binary evolution as a possible formation pathway of fast-rotating chemically enriched red giants. Studying the element abundances of the systems in our catalogue may uncover the properties and significance of these mechanisms in a different range of separations and binary evolution stages.

\begin{figure}
    \centering
    \includegraphics[width=\columnwidth]{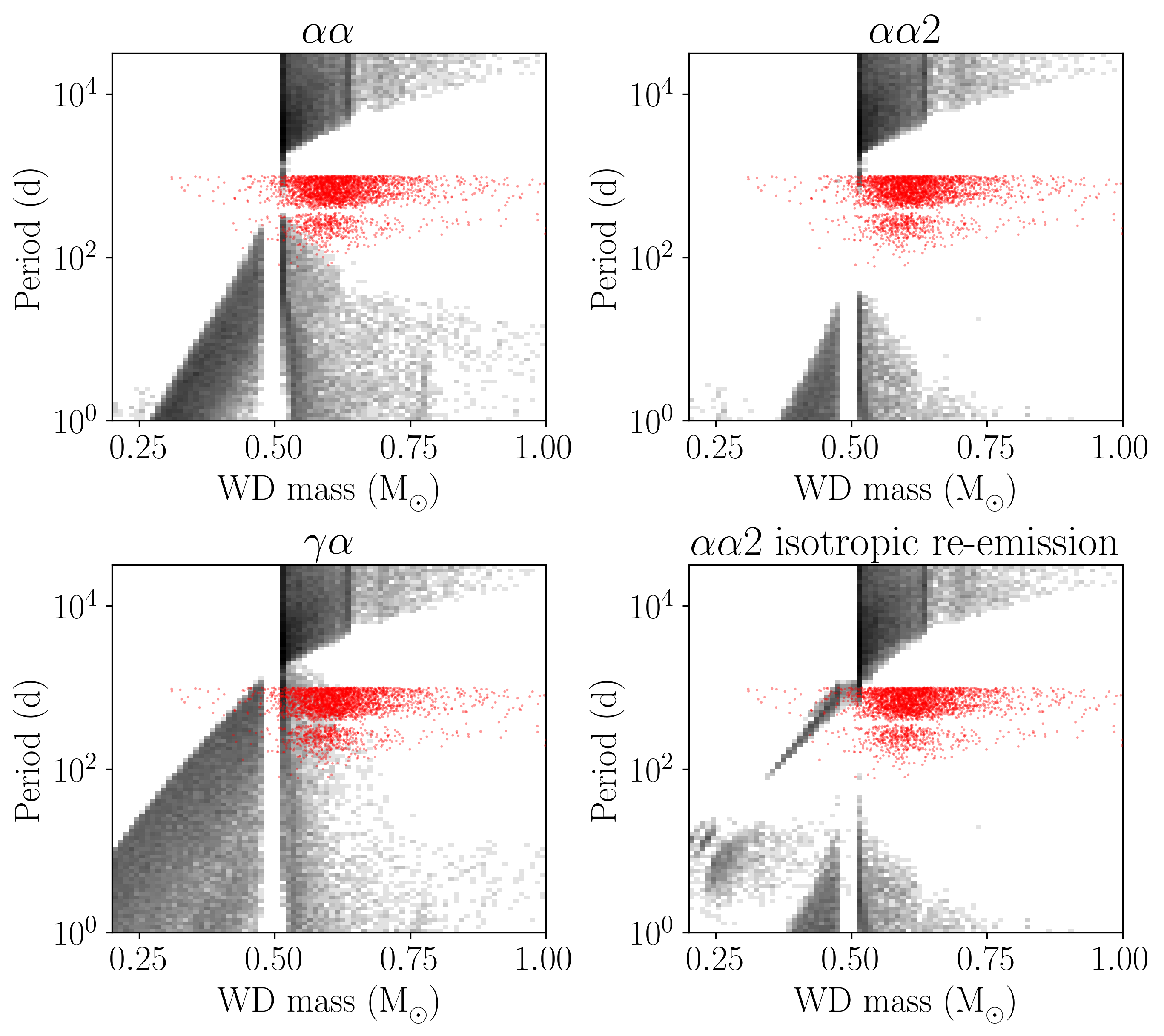}
    \caption{A comparison to a synthetic population of MS+WD binaries with $M_\textrm{MS}\leq1.2$\,\solarmass. Each panel shows the predicted period vs. WD mass distribution (coloured in logarithmic scale) assuming a different \texttt{SeBa} model (see text for details; \textit{clockwise from top left}): standard $\alpha\alpha$ model, standard $\alpha\alpha$ model with severe shrinkage in the common-envelope phase, the same as the previous but with isotropic re-emission, and the standard $\gamma\alpha$ model. Note that in the context of this study, only the treatment of the first mass transfer phase is relevant, and the two-phase model names are kept to ease the comparison with previous BPS studies.
    The \textit{Gaia} NCE sample is plotted as red dots.
    }
    \label{fig:BinaryEvolution}
\end{figure}

Fig.~\ref{fig:BinaryEvolution} compares the period-mass distribution of the observed NCE sample with the predicted distribution of various binary population synthesis (BPS) models. The synthetic populations shown here were generated using the \texttt{SeBa} BPS code \citep{PortegiesZwart_1996, Nelemans_2001, Toonen_2012}. The models differ in their treatment of the common-envelope phase. The process is parameterised using scalar factors governing the efficiency with which energy or angular momentum transfer unbind the envelope. These parameters are often denoted ``$\alpha$'' \citep{Paczynski_1976, Webbink_1984, Livio_1988} and ``$\gamma$'' \citep{Nelemans_2000, Nelemans_2005, vanDerSluys_2006}, respectively. 
The expulsion of the envelope shrinks the orbit of the binary in a process that can lead to the formation of a close double-WD system. These systems presumably experienced at least two phases of mass transfer, of which at least one occurred during a common-envelope phase. 

The BPS samples used here rely on two model families, $\alpha\alpha$ and $\gamma\alpha$ \citep{Toonen_2012}, defined based on the formalism used for each common-envelope phase. The $\alpha\alpha$ prescription, which is the standard model in most BPS codes (i.e. $\alpha\lambda=2$, where $\lambda$ is a parameter that depends on the structure of the donor star), has two additional variants: The $\alpha\alpha2$ model, capable of producing a more significant orbital shrinkage during the common-envelope phase (i.e. $\alpha\lambda=0.25$), was developed to match known populations of close MS+WD binaries \citep{Toonen2013,Zorotovic2010, Camacho2014,Scherbak2023}; The $\alpha\alpha2$ model with isotropic re-emission, in which mass is allowed to leave the system with the specific angular momentum of the accretor (instead of 2.5 times that of the orbit), is also capable of producing wider orbits. The resulting period versus WD mass distributions of all four prescriptions are presented in Fig.~\ref{fig:BinaryEvolution}, derived during the MS+WD evolutionary stage of the synthetic population. Note that in the context of this study only the treatment of the first mass transfer phase resulting in an MS+WD system is relevant, and the two-phase model names are kept to ease the comparison with previous BPS studies.

The observed NCE population is plotted as red dots in Fig.~\ref{fig:BinaryEvolution}. Although this population is affected by the selection criteria and observational biases, it is useful to consider whether current BPS codes are at all capable of producing it. We note that we only consider simulated MS+WD systems where the mass of the MS star is smaller than 1.2\,\solarmass, to reflect the selection in our observed sample. The $\alpha\alpha$ model and its variants have gaps around orbital periods of a few hundred days, where our observed NCE sample resides. Therefore, while some $\alpha\alpha$ variants might further improve by adjusting their parameters, this prescription seems less suitable for describing the observed NCE population. 
In contrast, the $\gamma\alpha$ model, which employs the $\gamma$ formalism for the first mass transfer event, succeeds in generating some MS+WD systems in part of the observed parameter range. This model was developed in order to match known populations of close double-WD binaries \citep{Nelemans_2000, Nelemans_2005, vanDerSluys_2006}.
We note, however, that all these models fail to reproduce MS+WD binaries with eccentricities larger than zero, in contrast with our observed sample.

If indeed the NCE sample represents a homogeneous MS+WD population, Fig.~\ref{fig:BinaryEvolution} suggests that some BPS prescriptions do not fully describe the binary population at the ${\sim}1$\,au separation range. This unprecedentedly large sample of MS+WD systems with intermediate orbital separations should help constrain and adjust these models to better fit the observations.

\subsubsection{WD initial to final mass relation}

Fig.~\ref{fig: nuv hrd} suggests we can estimate the cooling ages of many WDs in our sample based on their UV excess (but see the caveats discussion below). Since the WD masses were dynamically measured, estimating the binary's age would allow us to derive an empirical initial-to-final mass relation (IFMR) for these WDs. In most cases, these relations are obtained for single WDs in clusters \citep[e.g.][]{Cummings_2018}.

This sample allows for empirical estimation of the IFMR in binary systems with orbital separations potentially small enough to induce interaction during the late stages of the WD progenitor's evolution. However, determining the age of MS stars presents a challenge. Techniques like Gyrochronology \citep{bouma23} or abundance analysis may be helpful, yet their reliability, considering the configuration and evolutionary history of the binaries in our sample, is unclear (\citealt{silva23, gruner23}). An alternative approach would be associating MS+WD binaries with open clusters. Even though the numbers are relatively small, these systems provide a reliable age estimate for the WD and the binary as a whole.

\subsection{Caveats and limitations} 
As in the first paper, we rely heavily on the orbital solution's validity, the astrometric primary's mass estimate, and its classification as an MS star, as provided by \textit{Gaia}. Spurious astrometric measurements may lead to false detection of binary motion (see appendix E of \citealt{gaiaBH1}). Inaccurate estimates of the primary might affect the estimates regarding the nature of the faint companion, even if the orbit is valid (e.g., \citealt{el-badry24}).
However, the follow-up campaign for the relatively small \textit{class-III} sample in \citetalias{shahaf23} required dozens of observations over many months. Scaling up an initiated spectroscopic campaign to dynamically validate the orbits of thousands of systems might be impractical. Alternative approaches, such as non-dynamical validation via detailed modelling of the spectral energy distribution or archival velocity measurements, may be useful. 

The infrared-excess classification scheme aims to identify hierarchal triples, not MS+WD binaries. As a result, the RCE sample is not purely comprised of triple systems. The classification probabilities and estimated \textit{B-I} colour excess values are provided in Table~\ref{tab: table1} to enable other studies of this population that may require different confidence levels.  Furthermore, as discussed in Section~\ref{sec:biases}, the samples presented in this work are biased; astrophysical conclusions drawn based on them must take these biases into account. 

Astrometric binaries with an MS primary more massive than ${\sim}1.2 \solarmass$ will appear as \textit{class-I} binaries even if their companion is a WD. This is not a trivial limitation, as detecting A-type stars with WD companions is beyond the reach of the triage method. A complementary approach, using interferometric measurements, was recently presented in a series of publications by \citet{Waisberg2023g, Waisberg2023h}. Their analysis enabled the detection of companions to A-type MS stars at orbital periods on the order of a few tens of years which, in some cases, were identified as WD candidates or close binaries  (also see \citealt{Waisberg2023a, Waisberg2023b, Waisberg2023c, Waisberg2023d, Waisberg2023e, Waisberg2023f, Waisberg2023i, Waisberg2023j, Waisberg2023k, Waisberg2023l, Waisberg2023m, Waisberg2023n}).

An astrophysical scenario not considered in this work is a triple system in which the secondary companion is a binary comprised of a close binary of a WD and a low-mass MS star. While the occurrence and evolutionary path leading to such systems in this separation range is unclear, empirical constraints on their occurrence might be of interest \citep{Shariat2023}. In the context of this work, such a system could, in principle, demonstrate excess emission both in short and long wavelengths.

We also note that UV excess by itself is not a definitive proof of a WD companion, as a low-mass chromospherically active companion can produce UV excess (e.g., 2MASS J06281844-7621467; see \citealt{Lagos2022}). We emphasise that the classification scheme in this work relied on the astrometric orbit. UV emission, if detected, was used to corroborate the astrometric conclusion. Nevertheless, the possibility of activity-induced UV emission should be considered when studying individual systems from this sample, for example, by examining their spectral properties, persistence, and variability. This is particularly important when inferring the IFMR, as the WD cooling ages may be biased by activity-induced excess UV emission.

\subsection{Prospects for future work}
The samples presented in this work offer an opportunity to probe the biases and selection effects in the \textit{Gaia} catalogue. The NCE sample represents a homogeneous population of binary systems: The light contribution of both components in these binaries can be reliably constrained, their orbits are expected to be fairly circularised, and they are found at various Galactic latitudes. One can, therefore, use it to derive  \textit{Gaia}'s selection function for binary orbits \citep[e.g.,][]{Rix2021, Cantat-Gaudin2023, Castro-Ginard2023}.

\begin{figure}
    \includegraphics[width=\columnwidth]{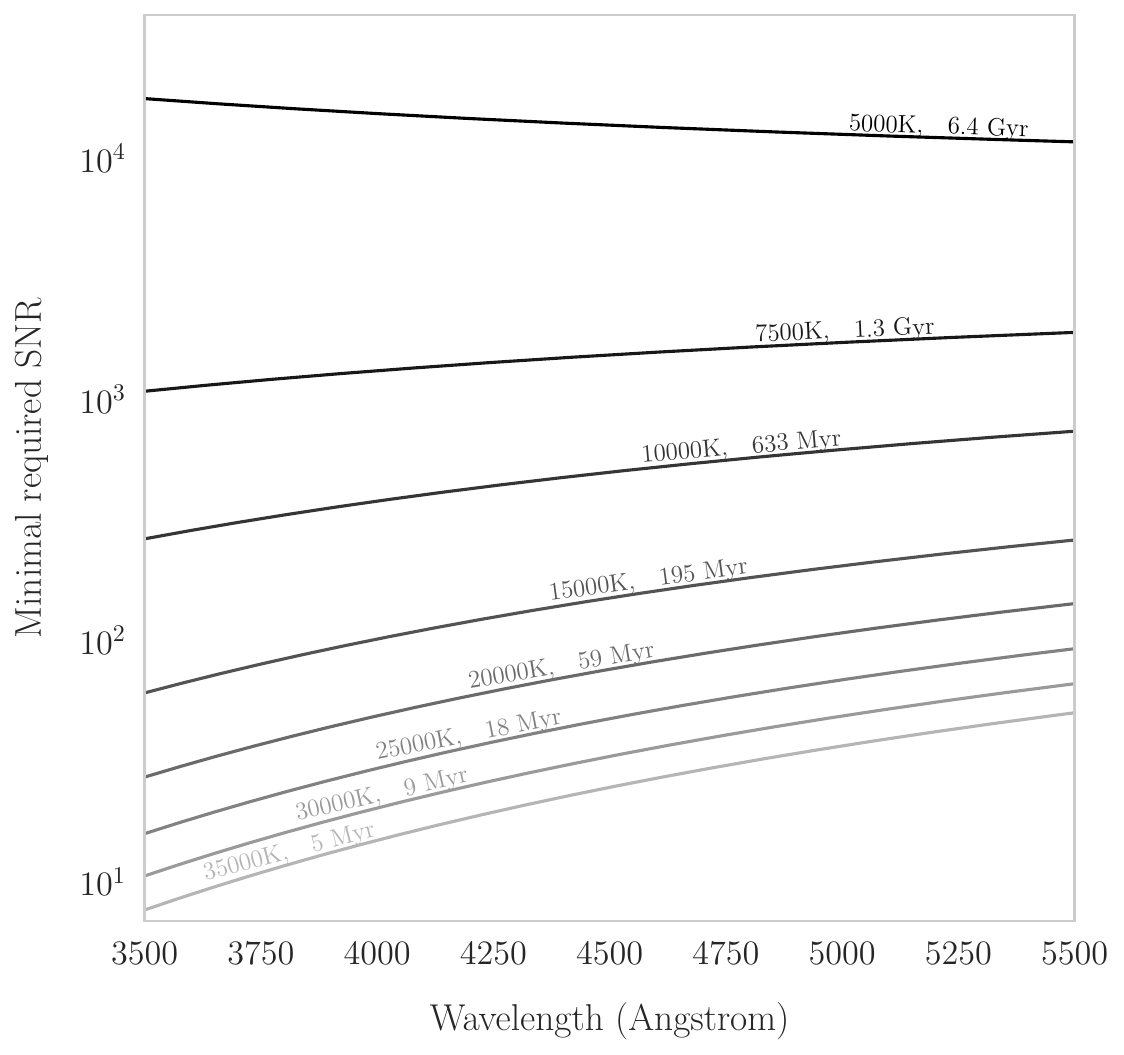}
        \caption{Minimal SNR required to detect a 0.6\,\solarmass\ WD companion to a 1\,\solarmass\ MS star. Each line corresponds to a different WD effective temperature (or cooling age). This calculation assumes black-body spectra for both components. The calculation is performed at a spectral resolution of $1\angstrom$.}
        \label{fig:MinSNR}
\end{figure}

For some science cases, validation of the orbits and classification validity may be required. Medium-resolution broadband spectroscopy offers a straightforward path to determine the nature of systems in our NCE sample. 
The spectra will allow us to characterise the primary star, leading to better constraints on its mass and metallicity. Excess emission at wavelengths $\leq450$\,nm ($\geq750$\,nm) might reveal the presence of a WD (late-type M-dwarf) companion. Fig.~\ref{fig:MinSNR} shows the signal-to-noise ratio (SNR) required to detect a WD at a given effective temperature (i.e. cooling age) versus wavelength assuming an MS primary of 1\,\solarmass. The component radii were estimated similarly to those of Fig.~\ref{fig:PhotometricExcess_G} (see Section~\ref{sec:biases}).

As the median magnitude of a target in our sample is relatively bright ($G=14.33$) and the measurements are not time-critical, a flexible program for telescopes operating in queue mode will be an effective follow-up strategy. The upcoming Son of X-Shooter (SoXS) spectrograph, expected to see first light in early 2024 on the European Southern Observatory (ESO) $3.6\,$m New Technology Telescope (NTT), is our case study for such a program. The spectrograph offers a resolution of $R{\sim}4\,000$ from $0.35$ to $2.1\upmu$m in a single exposure \citep{Schipani2020, Rubin2020}. Using the SoXS Exposure Time Calculator,\footnote{We assumed a black-body spectrum at $5\,700$K, airmass of $1.2$ and seeing conditions of $1\,$arcsec.} we find that a ${\sim}150$\,sec exposure is sufficient to achieve an SNR of $10$ at the short wavelength range of the spectrograph. At ${\sim}1\,\upmu$m a $25$\,sec exposure is required to reach the same SNR, similar to the exposure time required at ${\sim}0.4\,\upmu$m. 

Some studies, such as determining the WD IFMR, may require additional spaceborne UV measurements. For example, detailed studies of individual systems containing young WDs can utilise the Cosmic Origins Spectrograph (COS; \citealt{Osterman2011}) or the 
Advanced Camera for Surveys
 (ACS) on board the \textit{Hubble Space Telescope} (\textit{HST}) to determine their effective temperature, surface gravity and composition. On the other hand, the Wide-Field Camera 3 (WFC3; \citealt{Teplitz2013}) on the \textit{HST} can be used to determine open clusters and WD cooling ages efficiently. 

The samples in this work can provide insights into binary and stellar evolution and identify biases in the astrometric binary population. With the forthcoming fourth data release of \textit{Gaia}, our analysis can be extended to longer orbital periods. This will clarify the processes at the onset of mass transfer and their impact on binary evolution and the Galatic WD population.

\section*{Acknowledgements}
%===========================
We thank the anonymous referee for the valuable comments.
The discussions and interactions facilitated by the ``The Renaissance of Stellar Black-Hole Detections in The Local Group" workshop, hosted at the Lorenz Center in June 2023, significantly enriched this work. We thank the scientific organising committee and the Lorenz Center team for arranging the workshop. We thank Zephyr Penoyre, Simchon Faigler, and Dolev Bashi for their insightful comments and advice. The research of S.S. is supported by a Benoziyo prize postdoctoral fellowship. S.T. acknowledges support from the Netherlands Research Council NWO (VIDI 203.061 grants). K.E. was supported in part by NSF grant AST-2307232.

This work has made use of data from the European Space Agency (ESA) mission \href{https://www.cosmos.esa.int/gaia}{\it Gaia}, processed by the {\it Gaia} Data Processing and Analysis Consortium (\href{https://www.cosmos.esa.int/web/gaia/dpac/consortium}{DPAC}). Funding for the DPAC has been provided by national institutions, in particular the institutions participating in the {\it Gaia} Multilateral Agreement. 
This work made use of \href{https://github.com/naamach/stam}{\texttt{stam}}, a Stellar-Track-based Assignment of Mass package \citep{hallakoun21}; \href{https://artpop.readthedocs.io/en/latest/index.html}{\texttt{ArtPop}}, a Python package for synthesising stellar populations and simulating realistic images of stellar systems \citep{greco21}; the MIST isochrone grids \citep[][]{paxton11, paxton13, paxton15,choi16, dotter16}; The \href{https://github.com/SihaoCheng/WD_models}{WD models} package for WD photometry to physical parameters; \texttt{catsHTM}, a tool for fast accessing and cross-matching large astronomical catalogs \citep{soumagnac18}; \texttt{Astropy}, a community-developed core Python package for Astronomy \citep{Astropy_2013, Astropy_2018}; \texttt{matplotlib} \citep{Hunter_2007}; \texttt{uncertainties}\footnote{\href{http://pythonhosted.org/uncertainties/}{http://pythonhosted.org/uncertainties/}}, a \texttt{python} package for calculations with uncertainties by Eric O. Lebigot; \texttt{numpy} \citep{Numpy_2006, Numpy_2011}; and \texttt{scipy} \citep{2020SciPy-NMeth}.

%%%%%%%%%%%%%%%%%%%%%%%%%%%%%%%%%%%%%%%%%%%%%%%%%%
\section*{Data Availability}

All data underlying this research are publicly available.

%%%%%%%%%%%%%%%%%%%% REFERENCES %%%%%%%%%%%%%%%%%%

% The best way to enter references is to use BibTeX:

\bibliographystyle{mnras}
\bibliography{main}

%%%%%%%%%%%%%%%%%%%%%%%%%%%%%%%%%%%%%%%%%%%%%%%%%%

%%%%%%%%%%%%%%%%% APPENDICES %%%%%%%%%%%%%%%%%%%%%
\appendix
\section{Tables}

\begin{table*}
\begin{tabular}{rrrccrrrrrr}
\hline\hline
\multicolumn{1}{c}{Source ID} & \multicolumn{1}{c}{RA}    & \multicolumn{1}{c}{Dec}   & \multicolumn{1}{c}{$\varpi$ }       & \multicolumn{1}{c}{$M_1$ } & \multicolumn{1}{c}{$[{\rm Fe/H}]$} & \multicolumn{1}{c}{abs. V}  & \multicolumn{1}{c}{B--I}   & \multicolumn{1}{c}{$A_ {\rm V}$}   & \multicolumn{1}{c}{CE} & \multicolumn{1}{c}{Pr (red)} \\
          & \multicolumn{1}{c}{(Deg)} & \multicolumn{1}{c}{(Deg)} & \multicolumn{1}{c}{(mas)} &  \multicolumn{1}{c}{(${\rm M}_\odot$)}        &  \multicolumn{1}{c}{(dex)}  &   \multicolumn{1}{c}{(mag)} &  \multicolumn{1}{c}{(mag)}  &  \multicolumn{1}{c}{(mag)}   &  \multicolumn{1}{c}{(mag)} & \multicolumn{1}{c}{$(\%)$}   \\
 \hline
3334754343120640 & 48.893 & 5.471 & 3.01 & 0.6 & -0.945(61) & 8.227(55) & 2.779(11) & 0.493(96) & 0.25(21) & 89.43 \\
10378500708692608 & 49.532 & 7.405 & 6.73 & 0.7 & -0.543(37) & 7.888(76) & 3.082(10) & 0.027(29) & 0.37(25) & 96.56 \\
13499150931192576 & 53.145 & 11.734 & 1.91 & 0.8 & -0.383(77) & 6.79(12) & 2.407(12) & 0.740(20) & 0.25(19) & 90.83 \\
13966000991651456 & 47.822 & 8.848 & 5.39 & 0.7 & -0.480(35) & 7.500(59) & 2.7897(78) & 0.836(90) & 0.36(19) & 97.17 \\
14998957806126592 & 45.486 & 9.544 & 2.83 & 1.0 & -0.001(30) & 5.283(65) & 1.6924(43) & 0.466(37) & 0.02(12) & 57.57 \\
22172240385273088 & 42.461 & 10.675 & 1.77 & 0.9 & 0.346(31) & 5.40(16) & 1.8424(48) & 0.411(27) & 0.00(15) & 50.97 \\
28923104340937600 & 45.321 & 13.035 & 3.82 & 1.0 & -0.046(39) & 4.45(24) & 1.4547(47) & 0.384(20) & 0.08(15) & 72.67 \\
34377639791849984 & 46.833 & 16.147 & 4.02 & 1.2 & -0.824(91) & 3.62(15) & 1.2764(74) & 0.356(47) & 0.40(15) & 99.67 \\
37150234457357312 & 57.733 & 13.161 & 0.99 & 1.1 & -0.402(82) & 4.646(98) & 1.6268(53) & 0.630(47) & 0.23(13) & 95.79 \\
38662372182863104 & 58.068 & 13.413 & 1.00 & 1.0 & -0.255(86) & 4.593(96) & 1.7871(52) & 0.658(75) & 0.34(15) & 98.92 \\
\hline\hline
\end{tabular}
\caption{The cleaned sampled of non-\textit{class-I} objects in out sample. The Source ID, RA, Dec, and parallax were obtained from the \texttt{nss\_two\_body\_orbit} table, and the primary mass was obtained from \texttt{binary\_masses} table (Gaia Collaboration et al. 2022a). The absolute $V$-band magnitude and the $B$-$I$ colours were calculated using the \texttt{synthetic\_photometry\_gspc} table (Gaia Collaboration et al. 2022b), and appear in the table without extinction and reddening correction. The metallicity values used were selected as described in Section 2. CE and Pr (red) are the colour excess and the red excess probability, respectively. The full table is available in the supplemental information accompanying this publication.}
\label{tab: table1}
\end{table*}

\begin{table*}
\begin{tabular}{rrrrrrr}
\hline\hline
\multicolumn{1}{c}{Source ID} & \multicolumn{1}{c}{\textit{GALEX} ID} & \multicolumn{1}{c}{sep} & \multicolumn{1}{c}{NUV} & \multicolumn{1}{c}{FUV}  & \multicolumn{1}{c}{Excess} \\
 & & \multicolumn{1}{c}{(mas)} & \multicolumn{1}{c}{(mag)}  & \multicolumn{1}{c}{(mag)} & \multicolumn{1}{c}{(mag)} & \\ \hline
14998957806126592 & 6377073069995853037 & 0.94 & 19.080(60) &  & 0.6(1.1) &  \\
22172240385273088 & 6377213859023816129 & 0.68 & 21.53(33) &  & 1.4(1.1) &  \\
28923104340937600 & 6377073044226049205 & 1.25 & 17.063(28) &  & 0.33(87) &  \\
34377639791849984 & 6377073031343243419 & 0.74 & 14.787(11) & 20.18(25) & -0.18(60) &  \\
41408333753757056 & 6377073115093010630 & 1.98 & 22.46(46) &  & -6.5(1.1) & * \\
41954481793705984 & 6377073095765656506 & 2.27 & 20.98(24) &  & -5.5(1.8) & * \\
43892267960691840 & 6376967475003655762 & 0.4 & 18.490(60) &  & -0.2(1.3) &  \\
44082998867371776 & 6376967470708689546 & 0.6 & 21.71(38) &  & 0.8(1.1) &  \\
45281840202132224 & 6376967527617005236 & 0.32 & 18.355(32) &  & -0.1(1.1) &  \\
50144812630106240 & 6376861957322965342 & 0.7 & 20.14(14) & 19.81(16) & -4.7(1.8) & * \\
\hline \hline
\end{tabular}
\caption{UV data of 3\,672 binaries from our sample with an identified counterpart in \textit{GALEX}. The first three columns present the \textit{Gaia} and GALEX IDs and the angular separation in mili-arcsec between the two sources. The fourth and fifth columns provide the \textit{GALEX} apparent magnitudes without extinction correction. The sixth column provides the calculated excess absolute NUV magnitude (see text). An asterisk is used to indicate systems that were identified with excess UV emission. The full table is available in the supplemental information accompanying this publication.}
\label{tab: table2}
\end{table*}

%%%%%%%%%%%%%%%%%%%%%%%%%%%%%%%%%%%%%%%%%%%%%%%%%%

% Don't change these lines
\bsp	% typesetting comment
\label{lastpage}
\end{document}